\documentclass[sn-mathphys-ay]{sn-jnl}

\usepackage{graphicx}%
\usepackage{multirow}%
\usepackage{amsmath,amssymb,amsfonts}%
\usepackage{amsthm}%
\usepackage{empheq}
\usepackage{mathrsfs}%
\usepackage[title]{appendix}%
\usepackage{xcolor}%
\usepackage{textcomp}%
\usepackage{manyfoot}%
\usepackage{booktabs}%
\usepackage{algorithm}%
\usepackage{algorithmicx}%
\usepackage{algpseudocode}%
\usepackage{listings}%
\usepackage{lmodern}
\usepackage{anyfontsize}

\newtheorem{assumption}{Assumption}
\newcommand{\E}{\mathrm{E}}

\newcommand{\bbr}{\mathbb{R}}
\theoremstyle{thmstyleone}%
\newtheorem{theorem}{Theorem}%  meant for continuous numbers
\newtheorem{proposition}[theorem]{Proposition}% 

\theoremstyle{thmstyletwo}%
\newtheorem{example}{Example}%
\newtheorem{remark}{Remark}%

\theoremstyle{thmstylethree}%

\begin{document}

\title[Article Title]{Estimation of ratios of normalizing constants using stochastic approximation : the SARIS algorithm}


\author*[1]{\fnm{Tom} \sur{Gu\'edon}}\email{tom.guedon@inrae.fr}

\author[2]{\fnm{Charlotte} \sur{Baey}}\email{charlotte.baey@univ-lille.fr}

\author[1]{\fnm{Estelle} \sur{Kuhn}}\email{estelle.kuhn@inrae.fr}

\affil*[1]{\orgdiv{Universit\'e Paris-Saclay}, \orgname{INRAE, MaIAGE}, \orgaddress{\street{Domaine de Vilvert}, \city{Jouy-en-Josas}, \postcode{78350}, \country{France}}}

\affil[2]{\orgdiv{Laboratoire Paul Painlevé}, \orgname{Universit\'e de Lille}, \orgaddress{\street{Cit\'e scientifique}, \city{Villeneuve-d'Ascq}, \postcode{59655}, \country{France}}}


\abstract{Computing ratios of normalizing constants plays an important role in statistical modeling. Two important examples are hypothesis testing in latent variables models, and model comparison in Bayesian statistics. In both examples, the likelihood ratio and the Bayes factor are defined as the ratio of the normalizing constants of posterior distributions. We propose in this article a novel methodology that estimates this ratio using stochastic approximation principle. Our estimator is consistent and asymptotically Gaussian.  Its asymptotic variance is smaller than the one of the popular optimal bridge sampling estimator. Furthermore, it is much more robust to little overlap between the two unnormalized distributions considered. Thanks to its online definition, our procedure can be integrated in an estimation process in latent variables model, and therefore reduce the computational effort. The performances of the estimator are illustrated through a simulation study and compared to two other estimators : the ratio importance sampling and the optimal bridge sampling estimators.     }




\maketitle

\section{Introduction}

In statistical modeling, comparing models often hinges on estimating ratios of integrals, which frequently serve as normalizing constants of posterior distributions. For example in latent variables models, such ratios emerge when choosing between two nested models via the likelihood ratio test. Each marginal likelihood represents the normalizing constant of the posterior distribution density of the latent variables given the data. However, computing marginal likelihoods defined as integrals becomes infeasible when the relationship between observed data and latent variables is complex. Similarly, in Bayesian statistics, model selection can be performed by comparing evidences (or marginal likelihoods) between competing models through the Bayes factor. In this setting, marginal likelihoods serve as the normalizing constants of the posterior distribution of the parameters given the data. Therefore, being able to efficiently compute ratios of normalizing constants is of significant practical interest.

This topic has motivated many fields of applications such as phylogenetics \citep{lartillot2006computing}, astrophysics \citep{russel2018stepping}, psychology \citep{annis2019thermodynamic} or chemical physics \citep{shirts2008statistically}. To tackle this task one can either separately estimate the two likelihoods, or directly compute the ratio. As far as marginal likelihood estimation is concerned, several classical methods exist such as importance sampling \citep{robert1999monte}, the harmonic mean estimator \citep{newton1994approximate} or the generalized harmonic mean estimator \citep{gelfand1994bayesian}, using samples from the posterior distributions. It has been shown that those methods are often particular cases of estimators used to directly compute ratios of normalizing constants such as Bridge sampling \citep{meng1996simulating}  and ratio importance sampling \citep{chen1997monte}, both first introduced in the physics literature by respectively \cite{bennett1976efficient} and \cite{torrie1977nonphysical}. \cite{gronau2017tutorial} highlights that the Bridge sampling estimator is superior to both the importance sampling and the general harmonic mean estimators, as it is more robust to the choice of the proposal distribution.  However,  its performance deteriorates as the overlap between the two densities associated with the normalizing constants decreases. To circumvent this issue, refinements were proposed in \cite{meng2002warp} and \cite{wang2022warp}. They rely on a modification of the samples and their associated densities in order to increase the overlap between the densities. However, these refinements require some knowledge of the distributions under consideration, making the approach less attractive for practitioners. The optimal ratio importance sampling estimator presents the smallest asymptotic variance among the estimators mentioned here. However, the optimal scheme is not tractable in practice, which might explain why it is not often considered in the literature. Several other refined methods have been developed in the last decades, based on intermediate distributions that create a path between the two distributions. We can mention for example annealed importance sampling \citep{neal2001annealed}, sequential Monte Carlo \citep{del2006sequential}, path sampling \citep{gelman1998simulating} (or thermodynamic integration), stepping stone sampling \citep{xie2011improving} and generalized stepping stone sampling \citep{fan2011choosing}. These methods received a lot of interest and are presented and reviewed for instance in \cite{friel2012estimating} and more exhaustively in \cite{llorente2023marginal}. However, most of these schemes can be seen as refinements of elementary methods to compute ratios of normalizing constants presented in \cite{chen1997monte}. For example, path sampling is an extension of Bridge sampling \citep{gelman1998simulating}, and stepping stone sampling is an extension of the pre-umbrella identity using intermediate power posteriors distributions. 

We propose here a new approach to calculate ratios of normalizing constants based on a stochastic approximation algorithm. Our procedure is also related to ratio importance sampling, and is therefore called Stochastic Approximation of Ratio Importance Sampling (SARIS). It benefits from the different refinements available in these two fields. After showing that estimating a ratio of normalizing constants is equivalent to finding the root of a function defined as an expectation, we develop an iterative stochastic approximation scheme to compute this root. We show that the sequence generated by the proposed algorithm converges almost surely towards the targeted ratio.
The main advantage of our approach is that, thanks to its iterative construction, there is no need to fix the computational effort ahead since the procedure can be stopped once a convergence criterion has been reached.  We also express the optimal proposal distribution in terms of asymptotic variance using convergence results from stochastic approximation theory. Moreover our method allows to reach the same asymptotic variance as the theoretical one of the optimal ratio importance sampling estimator. 

This paper is organized as follows: the next section describes the context and the objective and gives a quick review of existing methods. Section \ref{sec:saris} presents the proposed SARIS procedures and studies their theoretical properties. Section \ref{seq:compmixt} is dedicated to algorithms using solely samples from both distributions involved in the targeted ratio. An extension for the simultaneous estimation of model parameters and likelihood ratio test statistic in latent variables models is also proposed. Section \ref{sec:numerical} is dedicated to numerical experiments and practical guidelines. We conclude and discuss the perspectives in Section \ref{sec:disc}.  The proofs are postponed to the appendix.

\section{Ratios of normalizing constants}\label{sec:intro}
In this section we introduce ratios of normalizing constants, the notations and their uses in statistics. We illustrate and motivate our purpose through two concrete examples: the likelihood ratio test statistic in latent variables models and the Bayes factor in Bayesian statistics. 

\subsection{Statistical setting and objective}\label{sec:context}
Let $d$ be a positive integer and $\mu$ a $\sigma$-finite positive Borel measure on a subspace $\mathcal{Z}$ of $\mathbb{R}^d$. Assume that $f_0$ and $f_1$ are two positive integrable Borel functions on $\mathcal{Z}$ such that $c_0 =  \int_{\mathcal{Z}}f_0(z)\mu(dz)>0$ and $c_1=  \int_{\mathcal{Z}}f_1(z)\mu(dz)>0$.
We assume that these normalizing constants   $c_0$ and $c_1$  are unknown and introduce  the two probability densities with respect to $\mu$ denoted by $p_0$ and $p_1$ defined  for $i \in \{0,1\}$ and for  all  $z$ in $\mathcal{Z}$ by 

\[ p_i(z) = \frac{f_i(z) }{c_i}.\]

%We consider two probability densities denoted by $p_0$ and $p_1$ defined on a subspace $\mathcal{Z}$ of $\mathbb{R}^d$ for some positive integer $d$. We assume that these densities are , known up to  normalizing constants denoted by  $c_0$ and $c_1$ respectively, such that for $i \in \{0,1\}$, for  all  $z$ in $\mathcal{Z}$: 
%
%\[ p_i(z) = \frac{f_i(z) }{c_i} \] 
%where $f_i$ is some known non-negative unnormalized density defined on $\mathcal{Z}$. 

The objective is to estimate the  ratio $r^*$ of normalizing constants defined as: 

\begin{equation}\label{eq:ratio}
    r^* = \frac{c_0}{c_1} = \frac{\int_{\mathcal{Z}}f_0(z)\mu(dz)}{\int_{\mathcal{Z}}f_1(z)\mu(dz)}.
\end{equation}

%\begin{remark}
%    We considered here the two densities defined on the same space $\mathcal{Z}$, but it does not impose anything on their support, it only requires that the densities are defined on spaces of the same dimension. 
%\end{remark}

%a repartir au fur et a mesure 
%When considering functions of two variables $(x,y)\mapsto m(x,y)$, we write $m(.,y)$ (respectively $m(x,.)$ the marginal functions $x\mapsto m(x,y)$ (respectively $y\mapsto m(x,y)$), for a given $x$ (resp. $y$). 

%We write $Z\sim\pi$ for $Z$ follows a distribution with density $\pi$. 
%Finally, $\underset{k\rightarrow + \infty}{\lim}x_k\overset{a.s.}{=}x^*$ and  ${x_k}\overset{d}{\underset{k\rightarrow+\infty}{\rightarrow}}\mathcal{D}$ stand respectively for \textit{the sequence $(x_k)_{k\geq0}$ converges almost surely to $x^*$} and \textit{the sequence $(x_k)_{k\geq0}$ converges in distribution to a random variable with distribution $\mathcal{D}$}.\\

We first motivate our contribution by two practical examples which require the computation of such ratios.  

\begin{example}[\textbf{computation of likelihood ratio test statistic in latent variables model}]\label{ex:lr}
Let us consider a general latent variables model where the observed variable is given by the random variable $Y$, taking values in $\mathcal{Y}$  and the latent variable by $Z$, taking values in $\mathcal{Z}$.  We denote by $y$ the observation of $Y$. We assume that the random vector $(Y,Z)$ follows a parametric distribution with density $f_\theta$  parameterized by $\theta\in\Theta$.  The objective is to test the hypotheses:
\[
  H_0: \quad \theta\in\Theta_0 \quad against\quad H_1:\quad \theta\in\Theta_1, 
 \]
where $\Theta_0  \subset \Theta_1  \subset \Theta$.
   A natural  popular test is the likelihood ratio test \citep{van2000asymptotic}  which statistic is defined by: 
\[
	LR = -2\log\left(\frac{L(\hat\theta_0;y)}{L(\hat\theta_1;y)}\right),
\]
and where the marginal likelihood $L(\theta;y) $ and the maximum likelihood estimates $\hat\theta_i$, for $i\in \{0,1\}$, are defined respectively by
  \[
  L(\theta;y) = \int_\mathcal{Z}f_\theta(y;z)\mu(dz)
 \] 
 and
\[
\hat\theta_i = \arg \underset{\theta\in\Theta_i}{\max}\quad L(\theta;y).
\]
Accurately estimating $LR$ is crucial as its value determines whether $H_0$ is rejected or not. 
\end{example}
%        Inference in latent variable models is usually performed by maximizing the marginal likelihood. Consider a random variable $Y$ with distribution known only jointly to a latent variable $Z\in\mathcal{Z}$. The complete likelihood $f_\theta$ is parameterized by $\theta\in\Theta$. The marginal likelihood of $\theta\in\Theta$, given data $y\in\mathcal{Y}$ is defined as:
%        \[L(\theta;y) = \int_\mathcal{Z}f_\theta(y;z)\mu(dz)\]
%    In most practical applications, inference is performed by stochastic methods such as stochastic gradient descent or a stochastic approximation version of the expectation maximization (EM) algorithm, that enables to avoid computing the intractable marginal likelihood. However, it is sometimes necessary to compute the marginal likelihood for a given $\theta$. This is the case in hypothesis testing contexts. To test: 
%    \[H_0: \quad \theta\in\Theta_0 \quad against\quad H_1:\quad \theta\in\Theta_1 \]
%    a natural and popular statistic of test is the likelihood ratio statistic (LR) defined as: 
%\[LR = -2\log\left(\frac{L(\hat\theta_0;y)}{L(\hat\theta_1;y)}\right) \]
%where $\hat\theta_i = \arg \underset{\theta\in\Theta_i}{\max}\quad L(\theta;y) $ (for $i=0,1$).
%Accurately estimating $LR$ is crucial as its value determines whether $H_0$ is rejected or not. \\\\

\begin{example}[\textbf{computation of Bayes factor for Bayesian model choice}]\label{ex:bayes}
    In Bayesian statistics,  the parameter $\theta\in\Theta$ is considered as a random variable with a known prior distribution $p(\theta)$. The posterior distribution of $\theta$ given a dataset $D$ is defined as the product of the prior density and the likelihood of the model. To compare two models $M_1$ and $M_2$ and choose the one that better fits the data, the Bayes factor $B_{12}$  \citep{gelfand1994bayesian} is a powerful tool.   It is defined as: 
    \[B_{12}=\frac{p(D \mid M_1)}{p(D\mid M_2)}\]
    where $p(D\mid M_i) = \int_\Theta p(D\mid\theta,M_1)p(\theta)d\theta$ is the marginal likelihood of model $M_i$, for $i \in \{1,2\}$.
\end{example}

\subsection{State of the art}\label{sec:stateoftheart}
When the ratio \eqref{eq:ratio} has no explicit expression, its computation can be performed  by  evaluating  separately the numerator and  the denominator. This can be done, for instance, using the harmonic mean estimator of \cite{newton1994approximate}. In a Bayesian context, this estimator uses draws from the posterior distribution to compute the inverse of the marginal likelihood. However, it is known to overestimate the marginal likelihood and can have infinite variance. Another solution is to use importance sampling \citep{robert1999monte}, which requires the introduction of a proposal distribution close to the integrand. However, it may be challenging to build such a proposal in complex settings. Furthermore, importance sampling is very sensitive to a misspecification of the proposal with respect to the density of interest \citep{gronau2017tutorial}. These methods are specific cases of more general ones that aim at estimating ratios of normalizing constants. 

In this section we focus on two existing methods that will serve as comparison for the proposed methodology: i) the Bridge sampling which is particularly popular \citep{meng2002warp,fruhwirth2004estimating,gronau2017tutorial,gronau2017bridgesampling}, and ii) the ratio importance sampling \citep{chen1997monte} which is strongly linked to our approach.
 For a more precise and exhaustive review we refer the reader to \citet[sections 4.1 and 4.2]{llorente2023marginal}, and to \cite{chen1997monte}. 
In the sequel, we denote by $Z$ any random variable defined on a probability space $(\Omega, \mathcal{A},P)$ taking values in  $\mathcal{Z}$, and by $\E_i$ the expectation with respect to density $p_i$ for $i\in\{0,1\}$.

\paragraph{Bridge sampling} First introduced in \cite{bennett1976efficient} and later reintroduced in \cite{meng1996simulating}, the Bridge sampling is based on the following identity: 
\begin{align*}
 r^* = \frac{\E_1\left[ f_0(Z)\alpha(Z)\right]}{\E_0\left[ f_1(Z)\alpha(Z)\right]},
\end{align*}
where $\alpha$ is a non-negative function defined on $\mathcal{Z}$ verifying $0<\int_\mathcal{Z}\alpha(z)p_0(z)p_1(z)\mu(dz)<+\infty$.  Let $K$ be a fixed positive integer. Then the Bridge sampling estimator of $r^*$ is obtained using two $K$-samples  $(Z^0_k)_{1 \leq k\leq K}$ and $(Z^1_k)_{1 \leq k\leq K}$ from $p_0$ and $p_1$ respectively as follows: 
\begin{equation}\label{eq:bs_est}
    \hat{r}_K^{BS} = \frac{\sum_{k=1}^{K} f_0(Z^1_k)\alpha(Z^1_k) }{\sum_{k=1}^{K} f_1(Z^0_k)\alpha(Z^0_k)}.
\end{equation}

This approach is particularly popular since in most statistical contexts it is straightforward to apply once a first inference step has been performed. This is the case for instance in the two examples introduced in the previous section.
In the first example dealing with hypotheses testing in latent variables models, $p_i$ is the posterior distribution of the latent variables given the data under hypothesis $H_i$. %In these models, the estimation process often requires draws from these posterior distributions (see also section \eqref{sec:joint} for more details). 
In the second example of Bayesian model choice, $p_i$ is the posterior distribution of the parameter given the data under model $M_i$.  In both contexts, sampling from these distributions is part of the entire estimation process, therefore no additional work is required regardless of the complexity of the distributions.%, which is part of Bridge sampling popularity. 

\cite{meng1996simulating} showed that the optimal choice of $\alpha$, that minimizes the mean square error of the estimator and its asymptotic variance, is given by:
\begin{equation}
   \alpha^{opt}_{bridge}(z) \propto \frac{1}{p_1(z) + p_0(z)}\propto \frac{1}{r^*f_1(z)+f_0(z)}.
\end{equation}
It reaches the following optimal normalized asymptotic variance: 
\begin{equation}\label{eq:var_opt_bs}
    V^{opt}_{bridge} = 4{r^*}^2\left[\left(\int_\mathcal{Z}\frac{2p_0(z)p_1(z)}{p_0(z)+p_1(z)}\mu(dz)\right)^{-1}-1\right].
\end{equation}

As $\alpha^{opt}_{bridge}$ depends on the unknown ratio $r^*$, it is not possible to use it directly in practice.  Therefore  the authors propose an iterative scheme to reach the optimal asymptotic variance. Starting from an initial guess $\hat{r}_K^{(0)}$, and using the two $K$-samples  $(Z^0_k)_{1 \leq k\leq K}$ and $(Z^1_k)_{1 \leq k\leq K}$ from $p_0$ and $p_1$ defined above, we get:
\begin{equation}\label{eq:optbridge}
    \hat{r}_K^{(t+1)} =  \left(\sum_{k=1}^K \frac{ f_0(Z^1_k)}{  \hat{r}_K^{(t)}f_1(Z^1_k)+f_0(Z^1_k)}\right) \left/ \left(\sum_{k=1}^K \frac{ f_1(Z^0_k)}{  \hat{r}_K^{(t)}f_1(Z^0_k)+f_0(Z^0_k)}\right) \right. .%\qquad t=1,2,...
%    \frac{\sum_{k=1}^K \frac{ f_0(z^1_k)}{  \hat{r}^{(t)}f_1(z^1_k)+f_0(z^1_k)}}{\sum_{k=1}^K \frac{ f_1(z^0_k)}{  \hat{r}^{(t)}f_1(z^0_k)+f_0(z^0_k)}}%\quad t=1,2,...
\end{equation}
As $t$ grows to infinity this estimator converges towards $\tilde{r}_K^{BS}$ defined as: 
\[\tilde{r}_K^{BS} = \left(\sum_{k=1}^K \frac{ f_0(Z^1_k)}{  \tilde{r}_K^{BS}f_1(Z^1_k)+f_0(Z^1_k)}\right) \left/ \left(\sum_{k=1}^K \frac{ f_1(Z^0_k)}{  \tilde{r}_K^{BS}f_1(Z^0_k)+f_0(Z^0_k)}\right)\right.\] 
%\frac{\sum_{k=1}^K \frac{ f_0(z^1_k)}{  \tilde{r}_Kf_1(z^1_k)+f_0(z^1_k)}}{\sum_{k=1}^K \frac{ f_1(z^0_k)}{  \tilde{r}_Kf_1(z^0_k)+f_0(z^0_k)}}\] 
which can be rewritten as: 
\[
\sum_{k=1}^K \frac{ f_0(Z^1_k)}{  \tilde{r}_K^{BS}f_1(Z^1_k)+f_0(Z^1_k)} -\sum_{k=1}^K \frac{\tilde{r}_K^{BS}   f_1(Z^0_k)}{  \tilde{r}_K^{BS} f_1(Z^0_k)+f_0(Z^0_k)} =0.
\]
This last equation shows that the optimal Bridge sampling estimator can also be defined as the root of a function. Note that the solution to this equation nullifies the score function of Geyer's likelihood described in \cite{geyer1994estimating}, leading to the reverse logistic regression estimator.  

Even if Bridge sampling is more robust than other methods as mentioned above, it still suffers from a too small overlap between distributions $p_0$ and $p_1 $. Indeed, when this overlap vanishes, the optimal variance grows to infinity. In such cases, more refined methods have been developed (see \cite{meng2002warp,wang2022warp}) that modify the two distributions considered without changing the normalizing constant. However, these methods are more involved since they require some additional effort to work properly.

\paragraph{Ratio importance sampling} 
The ratio importance sampling (RIS) estimator was first introduced in the physics literature in \cite{torrie1977nonphysical} as umbrella sampling and then rediscovered in \cite{chen1997monte}. It generalizes the importance sampling estimator to compute a ratio of normalizing constants. Considering a positive density function $\pi$ on $\mathcal{Z}$, dominated by $p_0$ and $p_1$, equation \eqref{eq:ratio} can be written as follows: 

\begin{equation}\label{eq:ris}
    r^* = \frac{\int_\mathcal{Z}f_0(z)\mu(dz)}{\int_\mathcal{Z}f_1(z)\mu(dz)} = \frac{\E_\pi\left[\frac{f_0(Z)}{\pi(Z)}\right]}{\E_\pi\left[\frac{f_1(Z)}{\pi(Z)}\right]}.
\end{equation}

Equation \eqref{eq:ris} is called the \textit{ratio importance sampling identity}, which will also be the basis of the methodology proposed in this paper. The ratio importance sampling estimator can be obtained using a $2K-$sample $(Z_k)_{1\leq k\leq 2K}$ from $\pi$ as follows:
\begin{equation}\label{eq:ris_est}
\hat{r}_K^{RIS} = \frac{\sum_{k=1}^{2K} f_0(Z_k)/ \pi(Z_k) }{\sum_{k=1}^{2K} f_1(Z_k)/\pi(Z_k)}.
\end{equation}

\begin{remark}
    Contrary to the Bridge sampling estimator that uses two samples, the Ratio Importance sampling estimator only requires one. That is why it is presented here using a sample of size $2K$. 
\end{remark}

Note that identity \eqref{eq:ris} is very interesting and very general. For example, by taking $\pi = p_1 $ it leads to $ r^* = \E_1\left[\frac{f_0(Z)}{f_1(Z)}\right] $ which gives an unbiased estimator of $r^*$. However, this estimator does not reach the optimal asymptotic variance. Indeed, %with the more general case presented in \eqref{eq:ris}, 
\cite{chen1997monte} showed that the optimal proposal density that minimizes the asymptotic variance of the estimator is given by:
\begin{equation}\label{eq:pioptiris}
%    z\mapsto\quad
    \pi^{opt}_{ris}(z)\propto |p_1(z)-p_0(z)|\propto|f_1(z)r^*-f_0(z)|.
\end{equation}
It reaches the following optimal asymptotic variance: 
\begin{equation}\label{eq:voptris}
    V_{ris}^{opt} = {r^*}^2 \left(\int_\mathcal{Z}|p_1(z)-p_0(z)|\mu(dz)\right)^2.
\end{equation}

\citet[Theorem 3.3]{chen1997monte} showed that the variance of the optimal ratio importance sampling estimator is smaller than the variance of the optimal Bridge sampling estimator. Furthermore, when the overlap between $p_0$ and $p_1$ goes to $0$, the optimal variance $V_{ris}^{opt}$ converges to $4r^*$ which is bounded.  This is a clear advantage compared to the previous methodology. However, contrary to the Bridge sampling setting, there is no straightforward procedure to approximate the optimal scheme and reach the optimal asymptotic variance. \cite{chen1997monte} suggested a two-stage approach consisting in building a first consistent estimator $\hat{r}$ of $r^*$ based on a chosen method, and then use it as a plug-in estimator in \eqref{eq:pioptiris}. Since the first step can be difficult to achieve, the optimal scheme can be difficult to implement in practice. This might partly explain why RIS is not as popular as Bridge sampling.

%To this extent , it is of interest to develop a new method that overcomes the issue of sensitivity to little overlap of Bridge sampling and applicability of ratio importance sampling. 

\section{Stochastic approximation procedures to compute ratio of normalizing constants}\label{sec:saris}
Given the limitations raised by the two approaches presented in the previous section, namely the sensitivity to a small overlap of the two distributions $p_0$ and $p_1$ for Bridge sampling, and the little practical applicability for RIS, there is a need for a new method that could address these issues.

In this section we propose an approach based on stochastic approximation principles.  Our procedure convert the ratio importance sampling identity \eqref{eq:ris} into a root finding program, which brings several advantages. First, and contrary to usual Monte Carlo computation, there is no need to fix the sample size ahead of the procedure. Second, the stochastic approximation framework enables the use of sampling distributions that depend on the current estimate of the unknown ratio $r^*$, circumventing the main obstacle of RIS, but still enjoying its good theoretical properties. Indeed, the obtained sequence of estimates is almost surely convergent and asymptotically Gaussian. It reaches the same asymptotic variance as the one of the optimal RIS estimator, with an applicable scheme. As a consequence, it is much more robust to little overlap between $p_0$ and $p_1$, avoiding the major drawback of Bridge sampling. 

\subsection{Description of the SARIS algorithm}\label{sec:desc}
%We describe here the proposed methodology in detail. 
Let $\pi$  be  a positive density function on $\mathcal{Z}$. Starting from equation \eqref{eq:ratio}, rewritten as:

\begin{equation}\label{eq:sa1}
   \int_\mathcal{Z} (f_0(z)-r^*f_1(z))\mu(dz)= 0,
\end{equation}
we can write:
\begin{equation}\label{eq:sa2}
    \E_{\pi}\left[\frac{f_0(Z)-r^*f_1(Z)}{\pi (Z)} \right]= 0,
\end{equation}
where $\E_\pi$ stands for the expectation with respect to the density $\pi$.  

Calculating $r^*$ is now equivalent to finding the root of a function defined as an expectation, and can therefore  be solved using stochastic approximation algorithms. Assuming that $r_0$ is given in $\mathbb{R}$ and that one can sample independent draws from $\pi$,  we thus consider the sequence $(r_k)_{k\geq0}$ of estimators of $r^*$, defined by  the following recursion   for every positive integer $k$: 
\begin{equation}\label{eq:saiid}
    r_{k+1} = r_k + \gamma_{k+1} \frac{f_0(Z_{k+1})-r_kf_1(Z_{k+1})}{\pi (Z_{k+1})},  \quad \text{with}  \ Z_{k+1}\sim\pi,
\end{equation}
and where $(\gamma_k)_{k\geq0}$ is a sequence of positive decreasing step sizes. 

The main task in the construction of the sequence is the choice of the proposal distribution $\pi$. Most of the relevant choices for $\pi$ might depend on the true ratio $r^*$.  For example the optimal choice for the Bridge sampling involves the quantity  $p_0+ p_1$,  depending  on $r^*$  which can not be evaluated. This is also the main drawback to the use of optimal ratio importance sampling. Thanks to its iterative nature, our methodology allows to consider proposal distributions which might depend on $r^*$. More precisely, let us consider a positive density function $\pi_r$ on $\mathcal{Z}$ which depends on $r$. Equation \eqref{eq:sa2} can be written as: 

\begin{equation}\label{eq:saetoile}
        \E_{\pi_{r^*}} \left[\frac{f_0(Z)-r^*f_1(Z)}{\pi_{r^*} (Z)} \right]= 0.
\end{equation}

%From now on, consider  families of non-negative densities on $\mathcal{Z}$: $\left\{\pi(.;r);\quad r\in\bbr\right\}$. As an example,  defining $\pi(.;r)\propto f_0 + rf_1$, the mixture between $p_0$ and $p_1$ is $\pi(.;r^*)$. 

Such equations can be solved using the Robbins-Monro algorithm \citep{robbins1951stochastic} that is based on the following stochastic recursion defined for every positive integer $k$: 

\begin{equation}\label{eq:algoRM}
    r_{k+1} = r_k + \gamma_{k+1} \frac{f_0(Z_{k+1})-r_kf_1(Z_{k+1})}{\pi_{r_k} (Z_{k+1})},  \quad \text{with} \ Z_{k+1}\sim\pi_{r_k}
\end{equation}

The general algorithm is called SARIS (Stochastic Approximation Ratio Importance Sampling) and is summarized in Algorithm \ref{alg:saris}.

\begin{algorithm}[H]
\caption{SARIS algorithm}\label{alg:saris}
\begin{tabbing}
    \qquad \enspace Input:  $(\gamma_k)_{k\geq0}$, $r_0$, stopping criterion  \\
    \qquad \enspace  Until stopping criterion: \\
    \qquad \qquad Draw $Z_{k+1}$ from $ \pi_{r_k}$ \\
    \qquad \qquad Update $r_{k+1} = r_k + \gamma_k \frac{f_0(Z_{k+1})-r_kf_1(Z_{k+1})}{\pi_{r_k} (Z_{k+1})} $\\
    \qquad \qquad $k=k+1$\\
    \qquad \enspace Return $r_k$
    %\qquad \enspace Compute $\emv{\theta}=(\emv{\beta},\emv{\Lambda},\emv{\sigma})=\arg \underset{\theta\in\Theta}{\sup}l(\theta;Y_{1:\n})$ \\
    
\end{tabbing}
\end{algorithm}

\begin{remark}
The iterative structure of our procedure enables to introduce a stopping criterion.  This is not the case in many other methods, in particular for ratio importance sampling and Bridge sampling. Indeed in those two cases it is not possible to compute $r_{k+1}$ given $r_k$, therefore one should fix the sampling size at the beginning of the procedure. 
%With the SARIS procedure, one can use any estimator $\hat{r}$ as a starting point of the procedure.
\end{remark}

\begin{remark}
In most real-life applications, it is difficult to independently and exactly simulate from complex distributions.  Therefore, the simulation step in \eqref{eq:algoRM}  might be intractable. It is however possible to use the transition kernel of an ergodic Markov Chain having $\pi_r$ as invariant distribution. One common practical choice for such Markov Chain Monte Carlo (MCMC) sampling scheme is the Metropolis-Hastings or the Metropolis-within-Gibbs algorithm \citep{robert1999monte}. The recursive scheme \eqref{eq:algoRM} can therefore be generalized as follows for every positive integer $k$:

\begin{equation}\label{eq:samd}
    r_{k+1} = r_k + \gamma_{k+1} \frac{f_0(Z_{k+1})-r_kf_1(Z_{k+1})}{\pi_{r_k} (Z_{k+1})}  \qquad Z_{k+1}\sim\Pi_{r_k}(.;Z_k)
\end{equation}
where $\Pi_{r}(.,.)$ is a transition kernel of an ergodic  Markov chain with invariant distribution $\pi_r$.
\end{remark}

\begin{remark}\label{rem:1marg}
  We emphasize that the SARIS algorithm can also be used to compute a single marginal likelihood. If we know a density $p$ only up to a normalizing constant $c$, $p = f/c$, then by introducing a known normalized density $g$, the $SARIS$ algorithm can be used to compute the ratio $r^*=c$.
% Indeed when the supports of the two distributions are not defined on the same spaces, the methodology could be more complicated to apply \citep{chen1997estimating}. 
\end{remark}

\begin{remark}
Finally, the proposed procedure enables to directly estimate any strictly monotonous and invertible transformation $g$ of the ratio. Suppose that the objective is to compute $g(r^*)$ (for example $g=-2\log$ to obtain a likelihood ratio statistic). The recursive scheme \eqref{eq:algoRM} can be easily modified to estimate $g(r^*)$ with the sequence $(g_k)_{k\geq0}$ defined as follows:

\begin{equation}\label{eq:satransfo}
g_{k+1} = g_k + \gamma_{k+1} \frac{f_0(Z_{k+1})-g^{-1}(g_k)f_1(Z_{k+1})}{\pi_{g^{-1}(g_k)} (Z_{k+1})}  \qquad 
     Z_{k+1}\sim\pi_{g^{-1}(g_k)}
\end{equation}
Algorithm \ref{alg:saris} can be easily adapted using this recursion, only changing the updating rule. 
\end{remark}

\subsection{Theoretical property of the SARIS algorithm}\label{sec:theo}

In this section we study the theoretical convergence property of the sequence $(r_k)_{k\geq0}$ generated  by the SARIS procedure described in Algorithm \ref{alg:saris}.
%\begin{equation}\label{eq:algoRM}
%r_{k+1} = r_k + \gamma_{k+1}\frac{f_0(Z_{k+1})-f_1(Z_{k+1})r_k}{\pi(Z_{k+1};r_k)}\qquad Z_{k+1}\sim\pi(.;r_k)
%\end{equation}
%where for every $r\in\bbr$,  $ \pi(.;r)$ is a positive density function on $\mathcal{Z}$.
We emphasize that such a setting is less general than the recursion defined in \eqref{eq:samd}, however its theoretical study corresponds to the one of \cite{robbins1951stochastic}. Moreover it allows a more fair comparison with the Monte Carlo methods presented in section \ref{sec:stateoftheart}, for which the theory was established for independent and identically distributed sampling. 

%avec H
%In this section, the introduce the following notation for the increment, for a given density functions family $\{\pi(.;r); r\in \bbr\}$ we define: 
%\[(z,r) \mapsto H_\pi(z,r) = \frac{f_0(z)-rf_1(z)}{\pi(z;r)}\]
%We first state the regularity assumptions on $H_\pi$ 

We first state some regularity assumptions on the functions $f_0, f_1$ and on the densities $\{\pi_r ,  r\in\bbr\}$:
\begin{assumption}\label{ass:density}
The functions $f_0$ and $f_1$ are positive integrable and for every $z\in\mathcal{Z}$, $r\mapsto \pi_r(z)$ is continuous.
    \end{assumption}

    \begin{assumption}\label{ass:supincrement}
\[    \E_0\left(  \underset{r\in\bbr}{\sup}  \left| \frac{f_0(Z)-rf_1(Z)}{\pi_r(Z)}\right|\right)+    \E_1\left( \underset{r\in\bbr}{\sup}   \left| \frac{f_0(Z)-rf_1(Z)}{\pi_r(Z)}\right|\right) <+\infty.\]
% where $\E_i$ stands for expectation with respect to the density $p_i$ for $i \in \{0,1\}$. 
    \end{assumption}

    %avec H
%    \begin{assumption}\label{ass:density}
%\[\E_0\left[\underset{r\in\bbr}{\sup}\quad|H_\pi(z,r)|\right]+\E_1\left[\underset{r\in\bbr}{\sup}\quad|H_\pi(z,r)|\right] <+\infty\]
%    \end{assumption}
    
This assumption ensures the integrability of the main quantities involved in the algorithm.
     %
%\begin{remark}
%Assumption \eqref{ass:density} could be lighten by considering compact sets of $\bbr$, but it is not of interest in this paper  as every considered relevant choice of proposal densities families will imply that  $H_\pi(z,.)$ is bounded for every $z\in \mathcal{Z}$.    
%\end{remark}
%
We also state a common assumption on the sequence of step sizes $(\gamma_k)_{k\geq0}$.

\begin{assumption}\label{ass:stepsize2}
The sequence $(\gamma_k)_{k\geq0}$ is positive, decreasing and verifies $\sum_{k=0}^{+\infty}\gamma_k=+\infty$ and $\sum_{k=0}^{+\infty}\gamma_k^2<+\infty$.
\end{assumption}

We can now state the almost sure (a.s.) convergence of the sequence $(r_k)$.

\begin{proposition}\label{prop:wp1 iid}
    Considering the sequence $(r_k)_{k\geq0}$ generated by Algorithm \ref{alg:saris}, under Assumptions \ref{ass:density}, \ref{ass:supincrement} and \ref{ass:stepsize2}, we get:  
    \[\underset{k\rightarrow +\infty}{\lim}\quad r_k= r^* \quad \text{a.s.} \]
\end{proposition}

The proof is postponed to the appendix.
We now require a stronger regularity assumption on the functions $f_0, f_1,\pi_r$  to derive the asymptotic distribution of the sequence $(r_k)_{k\geq 0}$.

\begin{assumption}\label{ass:nonprop}
The functions $f_0$ and $f_1$ are not proportional to each other.
\end{assumption}

\begin{assumption}\label{ass:regtcl}
    There exists $\delta>0$ such that 
    \[  \underset{k\geq 0}{\sup}\quad \E_0\left( \left|\frac{f_0(Z)-r_kf_1(Z)}{\pi_{r_k}(Z)}\right|^{1+\delta}\right)+    \underset{k\geq 0}{\sup}\quad\E_1\left(  \left|\frac{f_0(Z)-r_kf_1(Z)}{\pi_{r_k}(Z)}\right|^{1+\delta}\right)<+\infty. \]
\end{assumption}

\begin{assumption}\label{ass:stepsizetcl}
There exists  $\frac{1}{2}<\epsilon<1$, $a>0$, $b>0$ such that the sequence of step sizes $(\gamma_k)$ is of the form  $\gamma_k=\frac{a}{b+k^\epsilon}$.
\end{assumption}

These integrability and step sizes assumptions are  classical ones to obtain asymptotic normality results for martingales.
%
%avecH
%\begin{assumption}\label{ass:reg tcl}
%    There exists $\delta>0$ such that 
%    \[\underset{k\geq 0}{\sup}\quad \left(\E_0\left[ |H_\pi(Z,r_k)|^{1+\delta}\right]+\E_1\left[ |H_\pi(Z,r_k)|^{1+\delta}\right]\right)<+\infty \]
%\end{assumption}
%
The next result states the asymptotic normality of the averaged sequence defined as $(r^{AV}_k)_{k\geq0} = \left(\frac{1}{k}\sum_{j=0}^kr_j\right)_{k\geq0}$.\\

\begin{proposition}\label{prop:tcl}
    Considering the sequence $(r_k)_{k\geq0}$ generated by Algorithm \ref{alg:saris} and its averaged version $(r^{AV}_k)_{k\geq0}$, under Assumptions \ref{ass:density}, \ref{ass:nonprop}, \ref{ass:regtcl} and \ref{ass:stepsizetcl},  we get: 
    \[\sqrt{k}\left(r^{AV}_k-r^*\right)\overset{d}{\underset{k\rightarrow+\infty}{\longrightarrow} }\mathcal{N}\left(0,V_{saris}(\pi_{r^*})\right)\]
    with $$ V_{saris}(\pi_{r^*})=\frac{1}{{ c_1}^{2}} \E_{r^* }\left[\left(\frac{f_0(Z)-r^*f_1(Z)}{\pi_{r^*}(Z)}\right)^2\right]$$
    where the expectation $\E_{r^* }$ is taken with respect to the density $\pi_{r^*}$. 
    Furthermore, the optimal proposal $\pi^{opt}_{saris}$ defined as the one which minimizes the asymptotic variance
 %   \[\pi^{opt}_{saris} = \arg\underset{\pi}{\min}\quad V_\pi\]
    is given as 
    $$ \pi^{opt}_{r^*} (z) \propto |p_1(z) -p_0(z)|,$$
      corresponding to the optimal variance: 
    \[V_{saris}^{opt} = {r^*}^2  \left(\int_{\mathcal{Z}}|p_1(z)-p_0(z)|\mu(dz)\right)^2.\]
\end{proposition}

The proof is postponed to the appendix and relies on similar arguments as the derivation of optimal importance function in importance sampling (see \cite{robert1999monte} for more details).  We achieve the same optimal variance and retrieve the optimal proposition density of ratio importance sampling. Moreover in \cite{chen1997monte} the authors show that this variance is smaller than the variance of the optimal  Bridge sampling estimator. 

\begin{remark}\label{rem:mcmc}
If sampling is done through the use of a Markov transition kernel, similar theoretical results can be obtained assuming additional   regularity conditions on the Markov kernel.  For further details we refer to  \cite{allassonniere2015convergent} and \cite{ fort2015central}.
% However, these results involve quantities that are no longer explicit, which make them not relevant for theoretical comparison with existing methods.  \\
\end{remark}

It is important to notice that the algorithm presented in this section is not always applicable. In particular, the analytical expression of the optimal proposal density $\pi^{opt}_{r^*}$ is unknown, and therefore the update rule of the sequence $r_k$ defined in Algorithm \ref{alg:saris} is not computable. The next section solves this issue.

\subsection{A practical extension of the SARIS algorithm}

Suppose that for all $z$,  $\pi_r(z)=\tilde\pi_r(z)/c(r) $ where the analytical expression of $\tilde{\pi}_r(z)$ is known, then equation \eqref{eq:sa2} is equivalent to: 
\begin{equation}\label{eq:satilde}
        \E_{\pi_{r^*}} \left[\frac{f_0(Z)-r^*f_1(Z)}{\tilde\pi_{r^*} (Z)} \right]= 0.
\end{equation}

Based on this identity, one can consider the following extension of Algorithm \ref{alg:saris} called SARIS-EXT, that uses the known unnormalised density $\tilde\pi_r$ : 
\begin{algorithm}[H]
\caption{SARIS-EXT algorithm}\label{alg:sarisext}
\begin{tabbing}
    \qquad \enspace Input:  $(\gamma_k)_{k\geq0}$, $r_0$, stopping criterion  \\
    \qquad \enspace  Until stopping criterion: \\
    \qquad \qquad Draw $Z_{k+1}$ from $ \pi_{r_k} \propto \tilde{\pi}_{r_k}$ \\
    \qquad \qquad Update $r_{k+1} = r_k + \gamma_k \frac{f_0(Z_{k+1})-r_kf_1(Z_{k+1})}{\tilde\pi_{r_k}(Z_{k+1})}$\\
    \qquad \qquad $k=k+1$\\
    \qquad \enspace Return $r_k$
    %\qquad \enspace Compute $\emv{\theta}=(\emv{\beta},\emv{\Lambda},\emv{\sigma})=\arg \underset{\theta\in\Theta}{\sup}l(\theta;Y_{1:\n})$   
\end{tabbing}
\end{algorithm}

We state mild additional conditions on $\tilde\pi_r$ similar to Assumptions \ref{ass:supincrement} and \ref{ass:regtcl}, to prove that the sequence $(r_k)_{k\geq 0}$ generated by Algorithm \ref{alg:sarisext} verifies the same properties as the one generated by Algorithm \ref{alg:saris}.

\begin{assumption}\label{ass:supincrementtilde}
    \[    \E_0\left(  \underset{r\in\bbr}{\sup}  \left| \frac{f_0(Z)-rf_1(Z)}{\tilde\pi_r(Z)}\right|\right)+    \E_1\left( \underset{r\in\bbr}{\sup}   \left| \frac{f_0(Z)-rf_1(Z)}{\tilde\pi_r(Z)}\right|\right) <+\infty.\]
\end{assumption}

\begin{proposition}\label{prop:sarisextcvps}
    Considering the sequence $(r_k)_{k\geq0}$ generated by Algorithm \ref{alg:sarisext}, under Assumptions \ref{ass:density}, \ref{ass:stepsize2} and \ref{ass:supincrementtilde}, we get:  
    \[\underset{k\rightarrow +\infty}{\lim}\quad r_k= r^* \quad \textit{a.s.} \]
\end{proposition}

\begin{assumption}\label{ass:regtcltilde}
    There exists $\delta>0$ such that 
    \[ \underset{k\geq 0}{\sup} \quad \E_0\left( \left|\frac{f_0(Z)-r_kf_1(Z)}{\tilde\pi_{r_k}(Z)}\right|^{1+\delta}\right)+   \underset{k\geq 0}{\sup}\quad  \E_1\left(  \left|\frac{f_0(Z)-r_kf_1(Z)}{\tilde\pi_{r_k}(Z)}\right|^{1+\delta}\right)<+\infty. \]
\end{assumption}

\begin{assumption}\label{ass:c}
    There exists a neighborhood $U$ of $r^*$ such that $r\mapsto h(r) = \frac{c_0-rc_1}{c(r)}$ is continuously differentiable and $h'(r^*)<0$.
\end{assumption}

\begin{assumption}\label{ass:piopt}
    For all $r\in\bbr$ the set $\mathcal{A}_r = \left\{ z\in\mathcal{Z},\quad f_0(z) = rf_1(z) \right\}$ satisfies $\mu(\mathcal{A}_r )=0$. 
\end{assumption}

\begin{remark}
     $i)$ Assumption \ref{ass:c} is required to apply a central limit theorem on the sequence $(r_k)_{k\geq 0}$. For the SARIS algorithm, $h(r) = c_0-rc_1$ which does not require regularity conditions, as it is affine. 
     $ii)$ 
     Assumption \ref{ass:piopt} is required to ensure that $c^{opt}(r)$ defined as the normalizing constant of $z \mapsto \tilde\pi^{opt}_{r}(z) = |f_0(z) - r f_1(z)| \propto \pi^{opt}_{r}(z)$ verifies Assumption \ref{ass:c}. The following proposition extends the theoretical results of the SARIS algorithm to the SARIS-EXT algorithm.
\end{remark}

\begin{proposition}\label{prop:sarisexttcl}
    Considering the sequence $(r_k)_{k\geq0}$ generated by Algorithm \ref{alg:sarisext} and its averaged version $(r^{AV}_k)_{k\geq0}$, under Assumptions \ref{ass:density}, \ref{ass:nonprop}, \ref{ass:stepsizetcl}, \ref{ass:regtcltilde} and \ref{ass:c},  we get: 
    \[\sqrt{k}\left(r^{AV}_k-r^*\right)\overset{d}{\underset{k\rightarrow+\infty}{\longrightarrow} }\mathcal{N}\left(0,V_{ext}(\tilde{\pi}_{r^*})\right)\]
    Moreover we have : $$ V_{ext}(\tilde{\pi}_{r^*})=V_{saris}(\pi_{r^*}).$$
    Furthermore, under Assumption \ref{ass:piopt}, an optimal unnormalised proposal $\tilde{\pi}^{opt}_{r}$ defined as one of which minimizes the asymptotic variance
    %   \[\pi^{opt}_{saris} = \arg\underset{\pi}{\min}\quad V_\pi\]
    is given as 
    $$  \tilde{\pi}^{opt}_{r^*}(z) = |f_0(z) -r^*f_1(z)| \propto \pi^{opt}_{r^*} (z)$$
        corresponding to the optimal variance: 
    \[V_{saris}^{opt} = {r^*}^2  \left(\int_{\mathcal{Z}}|p_1(z)-p_0(z)|\mu(dz)\right)^2.\]
\end{proposition}

This proposition shows that theoretically, the extended algorithm has the same asymptotic performances as the initial SARIS algorithm.

\section{Non optimal methods using draws from $p_0$ and $p_1$}\label{seq:compmixt}

\subsection{Estimating ratios of normalizing constants using only distributions $p_0$ and $p_1$}
It is interesting to compare methods that use draws solely from $p_0$ and $p_1$. As explained in Section \ref{sec:context} with the two examples, in latent variable models, these represent the posterior distributions of latent variables given data under two different hypotheses. In Bayesian inference, they denote the posterior distributions of parameters under two distinct models. In both contexts, draws from these distributions are essential for inference, making it convenient to consider proposal distributions based on them. Numerically, this allows for the reuse of already simulated samples. Practically, it simplifies the method by eliminating the need to build new samplers. 

A natural choice is a mixture of $p_0$ and $p_1$, similar to the approach proposed by \cite{chen1997monte} in Section 5, and closely related to bridge sampling. We define the proposal density based on this mixture as follows:
\begin{equation}\label{def:mixt}
    z\mapsto\pi_{r^*}^{mixt}(z) = \frac{1}{2}\left\{p_0(z)+p_1(z)\right\} \propto f_0(z)+r^*f_1(z) 
\end{equation}

As long as we know how to draw samples from $p_1$ and $p_0$, it is easy to sample from this mixture, by sampling uniformly randomly from one or the other distribution. Of course the analytical expression of $\pi_{r}^{mixt}$ is unknown, but the extended algorithm is applicable considering the unnormalised density $\tilde\pi_{r}^{mixt}(z) = f_0(z) + r f_1(z)$. The simulating step in Algorithm \ref{alg:sarisext} requires to sample from $\pi_{r_k}^{mixt}$, which verifies for every $z\in\mathcal{Z}$ :  
\begin{align}\label{eq:pimixt}
    \pi_{r_k}^{mixt}(z) &\propto f_0(z)+r_kf_1(z) \nonumber\\
    &\propto c_0p_0(z)+r_kc_1p_1(z) \nonumber\\
    &\propto r^*p_0(z)+r_kp_1(z) \nonumber\\
    \Longleftrightarrow \pi^{mixt}_{r_k} &= \left(1-\frac{r_k}{r_k+r^*}\right)p_0(z)+\frac{r_k}{r_k+r^*}p_1(z)
\end{align}
where the weights of the mixture depend on $r^*$. Therefore it is not possible to sample from $\pi_{r_k}^{mixt}$ using simply draws from $p_0$ and $p_1$. 
\begin{remark}
We emphasize that even if it is not possible to sample from $\pi_{r_k}^{mixt}$ using draws from $p_0$ and $p_1$, it is still possible to use a MCMC procedure to sample from $\pi_{r_k}^{mixt}\propto f_0 + r_kf_1$, but this loses the practical benefits of using separately the two distributions. This scheme is still illustrated in the simulation study. 
\end{remark}

However, as simulating from the mixture $\pi^{mixt}_{r^*}$ is possible, one can consider the following alternative recursive scheme : 
\begin{equation}\label{eq:samixt}
    r_{k+1} = r_k + \gamma_{k+1}\frac{f_0(Z_{k+1})-r_kf_1(Z_{k+1})}{\tilde\pi^{mixt}_{r_k}(Z_{k+1})},\qquad Z_{k+1}\sim\pi^{mixt}_{r^*}
\end{equation}

which defines a new estimation procedure that is summarized in Algorithm \ref{alg:sarismixt}, and called SARIS-MIXT. 

\begin{algorithm}[H]
    \caption{SARIS-MIXT algorithm}\label{alg:sarismixt}
    \begin{tabbing}
        \qquad \enspace Input:  $(\gamma_k)_{k\geq0}$, $r_0$, stopping criterion  \\
        \qquad \enspace  Until stopping criterion: \\
        \qquad \qquad Draw $Z_{k+1}$ from $ \pi^{mixt}_{r^*} = \frac{1}{2}\left(p_0+p_1\right)$ \\
        \qquad \qquad Update $r_{k+1} = r_k + \gamma_k \frac{f_0(Z_{k+1})-r_kf_1(Z_{k+1})}{f_0(Z_{k+1})+r_kf_1(Z_{k+1})}$\\
        \qquad \qquad $k=k+1$\\
        \qquad \enspace Return $r_k$
        %\qquad \enspace Compute $\emv{\theta}=(\emv{\beta},\emv{\Lambda},\emv{\sigma})=\arg \underset{\theta\in\Theta}{\sup}l(\theta;Y_{1:\n})$   
    \end{tabbing}
    \end{algorithm}

\begin{remark}
    Contrary to the SARIS-EXT algorithm, in this procedure the distribution used for the sampling step does not depend on the current ratio $r_k$.
\end{remark}
It can be shown that under mild conditions, the sequence generated by Algorithm \ref{alg:sarismixt} converges almost surely towards $r^*$ and is asymptotically Gaussian. The proof follows the same lines as for the two other SARIS algorithms, the main steps are given in the proof of the  next result. 

Unfortunately, the estimator obtained from the algorithm SARIS-MIXT presents a higher asymptotic variance than the one obtained using SARIS-EXT with $\tilde\pi^{mixt}_r$ as proposal,  $V_{ext}(\tilde\pi^{mixt}_{r^*})$. The following proposition formalizes this statement. 

\begin{proposition}\label{prop:samixtcomp}
    Let $V_{saris}^{mixt}$ be the asymptotic variance of the averaged sequence generated by the SARIS-MIXT Algorithm \ref{alg:sarismixt}. Let $\tilde\pi^{mixt}_r(z) = f_0(z)+rf_1(z)$ be the unnormalised mixture between $p_0$ and $p_1$. Let $\Psi$ be the quantity defined as:
    \[    \Psi = \int_\mathcal{Z}\frac{p_1(z)p_0(z)}{\frac{1}{2}\left\{p_1(z)+p_0(z)\right\}}dz\]
    Under Assumptions \ref{ass:density}, \ref{ass:nonprop}, \ref{ass:stepsizetcl}, we get :
    \[V_{ext}(\tilde\pi^{mixt}_{r^*}) = 4{r^*}^2\left(1-\Psi\right)=\Psi^2V_{saris}^{mixt}.\]
\end{proposition}

The quantity $\Psi$ can be seen as an overlap index between $p_0$ and $p_1$.  It is easy to show that $0\leq\Psi\leq 1$. %, it appears for example in the supplementary material of  \cite{walker2021new}.
If the two distributions have disjoint supports, then $\Psi=0$. If $p_0=p_1$ then $\Psi=1$.

This index $\Psi$ is very convenient to compare the asymptotic variances derived in Proposition \ref{prop:samixtcomp} with those of the optimal Bridge sampling estimator and ratio importance sampling estimator using $\pi^{mixt}_{r^*}$ as proposal. For a detailed description of the later called Bridge-like ratio importance sampling method (RIS-MIXT), we refer to \cite{chen1997monte}, section 5. Let $V_{ris}^{mixt}$ denote the asymptotic variance of the Bridge-like ratio importance sampling estimator. The following relationship exists between the various variances discussed in this paragraph:
\begin{equation}\label{eq:compmixt}
    V_{ext}(\tilde\pi^{mixt}_{r^*})=\Psi\times V^{opt}_{bridge }= \Psi^2\times  V^{mixt}_{ris} = \Psi^2\times  V^{mixt}_{saris}
\end{equation}
Note that the SARIS-MIXT estimator reaches the same asymptotic variance as the Bridge-like ratio importance sampling estimator, which is not surprising as they are both based on the same identity \eqref{eq:ris}. 

It is noticable that $V_{ext}(\tilde\pi^{mixt}_{r^*})$ is bounded, unlike the other variances that diverge as $\Psi$ approaches $0$. In fact, when $\Psi=0$, $\pi^{mixt}_r$ aligns with $\pi^{opt}_r$. Therefore, finding a way to sample from the distribution defined by \eqref{eq:pimixt} using only draws from $p_0$ and $p_1$ would be very beneficial. This would allow for variance reduction, as described in equation \eqref{eq:compmixt}, while still maintaining the simplicity of simulating only from $p_0$ and $p_1$. This result indicates that Bridge sampling is still the best method when using only draws from $p_0$ and $p_1$. However, when $\Psi$ is closed to one (which is the case with a lot of overlap between $p_0$ and $p_1$), these methods remain comparable. 

The next section explores the extension of the SARIS-MIXT algorithm to a joint procedure with parameter inference in latent variables models.

\subsection{A joint procedure for model parameter estimation  and LRT statistic computation in latent variables models}\label{sec:joint}

In this section, we extend the use of the SARIS estimator based on mixtures between $p_0$ and $p_1$ to the context of likelihood ratio test (LRT) in latent variables model and introduce a joint procedure for model parameter inference and LRT statistic computation.

%propose to integrate the SARIS algorithm to model parameter estimation procedures in latent variable models  to evaluate simultaneously parameter estimates and LRT statistic. 

Consider two random variables $Y$ on $\mathcal{Y}$ and $Z$ on $\mathcal{Z}$. Assume that the joint density of $(Y,Z)$ belongs to a parametric family $\{f_\theta, \theta\in\Theta\}$, with $\Theta\subset\bbr^p$ with $p$ a positive integer. We only observe a realization $y$ of $Y$, the random variable  $Z$ being unobserved. The maximum likelihood estimator $\hat{\theta}$ is defined as: 
\[\hat{\theta} = \arg \underset{\theta\in\Theta}{\max} \quad L(\theta;y)\]
where the marginal likelihood $L(\theta;y)$ is equal to the complete likelihood integrated over the latent variable: 
\[L(\theta;y) = \int_{\mathcal{Z}}f_\theta(y,z)\mu(dz)\]

The above integral is often untractable, which makes the optimization process difficult. To solve this issue, stochastic methods can be used. Two popular ones are the stochastic approximation expectation maximization (SAEM) algorithm \citep{delyon1999convergence,kuhn2004coupling} and the stochastic gradient descent (SGD) algorithm \citep{baey2023efficient}. Both methods require draws from the posterior distribution of the latent variables given the data whose density is denoted by $p_\theta$ in the sequel.
%defined as: 
%\[z\mapsto\quad p_\theta(z) = \frac{f_\theta(y,z)}{L(\theta;y)}\] 
%where the dependency of $p_\theta$ on $y$ is omitted. 

Both SAEM and SGD are iterative algorithms that can be summarized as follows, at each step $k>0$:

\begin{enumerate}
    \item Draw $Z_k$ from $p_{\theta_k}$.
    \item Update $\theta_k$ with a gradient step when using SGD or a maximization step when using SAEM.
\end{enumerate}

Consider now the context of Example \ref{ex:lr} where the objective is to test the hypotheses:
  \[H_0: \quad \theta\in\Theta_0 \quad against\quad H_1:\quad \theta\in\Theta_1 \]
where $\Theta_0  \subset \Theta_1  \subset \Theta$.
  The LRT statistic equals:
\[
LR = -2\log\left(\frac{L(\hat\theta_0;y)}{L(\hat\theta_1;y)}\right).
 \]

We propose to combine the estimation procedures for $\hat\theta_0$ and $\hat{\theta}_1$ with the computation of the marginal likelihood ratio $r^* = \exp(-LR/2)$, taking advantage of the computational effort of the inference task. The procedure is detailed in Algorithm \ref{alg:joint}.

%Consider two inference tasks on two different parameter spaces $\Theta_0\subset\Theta_1$, that aim at estimating the maximum likelihood estimators $\hat{\theta}_i$ $(i=0,1)$: 
%\begin{equation*}
%    \left\{\begin{array}{cc}
%         \hat\theta_0 = \arg \underset{\theta\in\Theta_0}{\max} \quad L(\theta;y)  \\
%           \hat\theta_1 = \arg \underset{\theta\in\Theta_1}{\max} \quad L(\theta;y)
%    \end{array}\right.
%\end{equation*}
%To compute the likelihood ratio:
%\[r^* = \frac{L(\hat\theta_0;y)}{L(\hat\theta_1;y)}\]
%we propose an algorithm to simultaneously estimate  $\hat\theta_0$, $\hat{\theta}_1$ and $r^*$.

\begin{algorithm}[H]
    \caption{Joint parameter and LRT statistic estimation in latent variables models}\label{alg:joint}
    \begin{tabbing}
\qquad \enspace Input: $z_{0,0}, z_{1,0}, \theta_{0,0}, \theta_{1,0}, r_0, (\gamma_k)_{k\geq0}$\\
\qquad \enspace $k=0$\\
\qquad \enspace Until convergence criterion: \\
\qquad \qquad Draw $z_{0,k+1}$ from $p_{\theta_{0,k}}$ and $z_{1,k+1}$ from $p_{\theta_{1,k}}$\\
\qquad\qquad Update $\theta_{0,k+1}$ and $\theta_{1,k+1}$ using a SGD or SAEM step\\
\qquad \qquad Draw $\tilde{z}_{k+1}$ from a uniform distribution on $\{z_{0,k+1},z_{1,k+1}\}$\\
 \qquad\qquad Update $r_{k+1}$ as: 
    \end{tabbing}
    \[r_{k+1} =r_k+ \gamma_k \frac{f_{\theta_{0,k+1}}(y,\tilde{z}_{k+1})-r_kf_{\theta_{1,k+1}}(y,\tilde{z}_{k+1})}{f_{\theta_{0,k+1}}(y,\tilde{z}_{k+1})+r_kf_{\theta_{1,k+1}}(y,\tilde{z}_{k+1})}\]

    \begin{tabbing}
     \qquad \qquad $k= k+1$   \\
     \qquad \enspace Return $r_k, \theta_{0,k}, \theta_{1,k}$
    \end{tabbing}
\end{algorithm}

%The proof of the convergence of the sequence $(r_k)_{\geq0}$ is beyond the scope of this paper and is left as a future investigation. 
\begin{remark}
When the two estimation processes can not be applied jointly, this procedure can be carried out post-estimation, provided the sequences $(\theta_{i,k}, z_{i,k})_{i=0,1, k\geq0}$ are kept in memory.
\end{remark}

\begin{remark}
If only one marginal likelihood is to be estimated, the procedure applies by introducing a proposal density $q$ (or a sequence of proposal densities $(q_k)_{k\geq 0}$) from which it is possible to sample from and to proceed as follows: 

\begin{enumerate}
    \item Draw $z_{k+1}$ from $p_{\theta_{k}}$ 
    \item Update $\theta_{k+1}$ with  SGD or SAEM step
    \item With probability $0.5$ define $\tilde{z}_{k+1} =z_{k+1} $, otherwise draw $\tilde{z}_{k+1}$ from $q_{k+1}(.)$
    \item Update $r_{k+1}$ as: 
    \[ r_{k+1} =r_k+ \gamma_k \frac{f_{\theta_{k+1}}(y,\tilde{z}_{k+1})-r_kq_{k+1}(\tilde{z}_{k+1})}{f_{\theta_{k+1}}(y,\tilde{z}_{k+1})+r_kq_{k+1}(\tilde{z}_{k+1})}   \]
\end{enumerate}

The R package \texttt{bridgesampling} \citep{gronau2017bridgesampling} proposes to use a Gaussian approximation as a second distribution when considering the computation of a single marginal likelihood. However, in our procedure prior knowledge on the distribution of interest (mean and variance for example) is not available as inference has not been performed yet. A possible approach to overcome this issue could be to use a Gaussian proposal with adaptive mean $m_k$ and variance $\sigma^2_k$, where $m_k$ and $\sigma^2_k$ are defined at each step $k$ as follows: 
\begin{align*}
    m_{k+1} & = m_k+\gamma_k(z_{k+1}-m_k)\\
    v_{k+1} & = v_k + \gamma_k(z_{k+1}^2-v_k) \\
    \sigma^2_{k+1} & = v_{k+1}-m_{k+1}^2
\end{align*}
\end{remark}

% The next section is dediated to some numerical experiment to investigate the performance of the proposed methodology.
 
\section{Numerical experiments and practical considerations}\label{sec:numerical}

This section is devoted to numerical experiments. We first illustrate the performances of the three SARIS estimators compared to the RIS estimator of \cite{chen1997estimating} and the optimal Bridge sampling estimator of \cite{meng1996simulating}. We then provide an example of the joint procedure introduced in Section \ref{sec:joint}.

\subsection{Simulation study in a one dimensional Gaussian setting}

We first illustrate our method with a one dimensional Gaussian setting. We consider two Gaussian distributions,  $f_0 = \phi$ and $f_1 =\phi(\cdot-\mu)$ where $\phi$ is the standardized Gaussian density and $\mu\in \mathbb{R}$. Since these densities are already normalized, we have $c_0=c_1=1$ and therefore $r^*=1$. 

In the simulation study,  we compare the performances of three SARIS estimators presented in this paper. These methods are compared with the optimal bridge sampling estimator (BRIDGE OPT) and the RIS estimator based on $\pi^{mixt}_{r^*}$ (RIS-MIXT). For the entire simulation study, the sampling steps are performed using one step of an adaptive Metropolis Hastings (MH) algorithm (see for example \citet[section 3]{roberts2009examples}) in order to stick to most real life applications, implemented manually.

\begin{figure}[htbp]
    \centering
    \begin{minipage}[b]{0.48\textwidth}
        \centering
        \includegraphics[width=\textwidth]{ 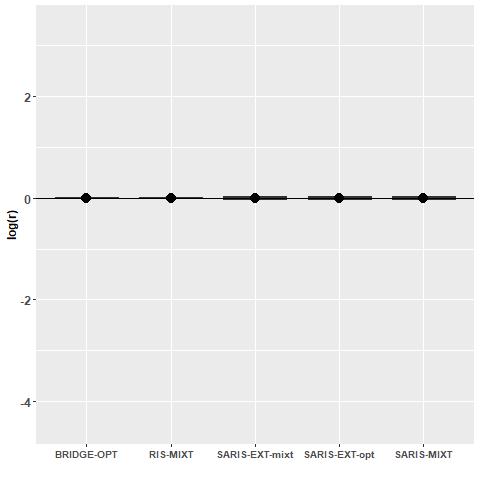}
    \end{minipage}
    \hfill
    \begin{minipage}[b]{0.48\textwidth}
        \centering
        \includegraphics[width=\textwidth]{ 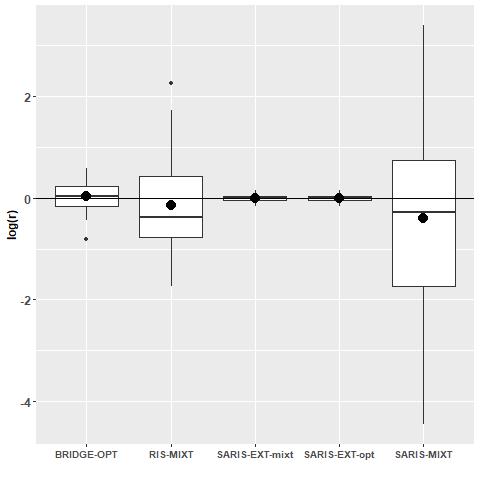}
    \end{minipage}
    \caption{Estimation of $\log r^*$ using the optimal bridge sampling estimator (BRIDGE-OPT), the bridge like ratio importance sampling (RIS-MIXT), SARIS-EXT-mixt, SARIS-EXT-opt and SARIS-MIXT. The black line represents the true value $log(r^*)=0$. The boxplots were computed on 50 repetitions of the experiments. The two graphs illustrate strong overlap between $p_0$ and $p_1$ (on the left, in the case $\mu=1$) and little overlap (on the right, in the case $\mu=5$)}\label{fig:compmu}
\end{figure}

The three SARIS estimators presented in Figure \ref{fig:compmu} are the following:

\paragraph{1) Optimal SARIS extended using $\tilde\pi^{opt}_r$ (SARIS-EXT opt)}
This is the estimator generated by Algorithm \ref{alg:sarisext} using $\tilde{\pi}_{r}^{opt}(z) = |f_0(z) - rf_1(z)|$. 
In this procedure, the increment $\frac{f_0(Z_{k+1})-r_kf_1(Z_{k+1})}{|f_0(Z_{k+1})-r_kf_1(Z_{k+1})|}\in\left\{0;1\right\} $ only takes two values. The drawback is that the increment has no intensity, i.e. it gives no indication on the order of magnitude of the descent step. However, it can still solve computational issues, in particular when the evaluation of the likelihood can be complicated. In order to use this method, one only need to know how to evaluate the unnormalised densities up to a non decreasing transformation, as only comparison of them is required, and not their evaluation.
%, and this proposal distribution enables to circumvent this issue by allowing to only evaluate the log likelihood.  Indeed we have for every $z\in\mathcal{Z}$ , and $r\in\bbr_+$: \begin{align*}
   % H_{\pi^{opt}}(z,r)&= sign\left(f_0(z)-rf_1(z)\right)\\
%    &= sign\left(\log(f_0(z))-\log(r)-\log(f_0(z))\right)
%\end{align*} 

\paragraph{2) SARIS using mixture between $p_0$ and $p_1$ (SARIS-MIXT)} 
%In many applications, simulating from the posterior distribution is part of the whole estimation process (Bayesian statistics, latent variables models). Therefore, it might be convenient to directly use draws from these distributions.

This distribution has already been discussed in Section \ref{seq:compmixt}. 
As long as we know how to draw samples from $p_1$ and $p_0$,  it is easy to sample from the mixture by randomly sampling from one or the other distribution with probability 1/2:1/2. This estimator corresponds to the one generated by Algorithm \ref{alg:sarismixt}.

%TROP DETAILLE ON SAIT FAIRE 
%In order to draw $Z$ from $\pi^{mixt}(.;r^*)$, one can proceed in two steps: 
%\begin{enumerate}
%    \item Draw $u$ uniformly over the interval $[0,1]$
%    \item If $u<0.5$, draw $Z$ from $p_0$, otherwise draw $Z$ from $p_1$
%\end{enumerate}

\paragraph{3) SARIS extended using $\tilde\pi^{mixt}_r = f_0(z) + rf_1(z)$ (SARIS-EXT-mixt)} 
This proposal is not of practical use, as it neither uses draws from $p_0$ and $p_1$ nor is an approximation of the optimal scheme. However, it is interesting to distinguish it from the previous one, as they are very similar. This distribution is also a mixture between $p_0$ and $p_1$ as explained in Section \ref{seq:compmixt}. Even if it is supposed to approximate $\pi^{mixt}_{r^*}$, it benefits from a significant gain in performance, as Proposition \ref{prop:samixtcomp} theoretically justifies it, and Figure \ref{fig:compmu} illustrates it. \\

Note that the three different proposal considered above lead to a bounded increment.
%: for every $z\in\mathcal{Z}$ and every $r\in\bbr_+$, $-1\leq  H_\pi(z,r) \leq 1$. 
This property is crucial both from a practical point of view, as it enhances the numerical stability of the procedure, and a theoretical one, as it guarantees Assumptions \ref{ass:supincrement} and \ref{ass:regtcl} to be verified.

\begin{remark}
    The optimality criterion considered in this article is given by the asymptotic variance of the \textit{exact} sampling scheme, which is most of the time intractable in practice. Indeed, in most real life applications sampling is performed through the use of transition kernels in MCMC algorithms. In this setting, even if central limit theorems may apply under regularity conditions, the asymptotic variance is in general not explicit and there is no guarantee that the proposal which minimizes this variance is the same as in the exact sampling case. Therefore it might still be of practical interest to consider other proposals. Some proposal distributions are worth mentioning, e.g. the geometric mean between $p_0$ and $p_1$ discussed in \cite{meng1996simulating} and \cite{chen1997estimating} for example, or an element of the $q$-path between $p_0$ and $p_1$ that generalizes the harmonic mean and geometric mean between the two distributions (see \cite{brekelmans2020annealed} for details).
\end{remark}

For each of the SARIS procedure the following step size is considered: 
\begin{equation}
\begin{cases}
          \gamma_k = 0.1 & \text{for }k<K_{heat}\\
          \gamma_k = \frac{0.1}{1+k^{2/3}} & \text{otherwise},
\end{cases}
\end{equation}
which is standard in the stochastic gradient descent literature. It verifies Assumption \ref{ass:stepsize2} and \ref{ass:stepsizetcl} while presenting a heating phase that enables a wider exploration of the parameter space at the beginning of the algorithm.

%\textit{ Both  estimators are not explicit and are defined implicitly.  Let $K$ a given positive integer.
%The Bridge like ratio importance sampling estimator $\hat{r}^{ris}_K$ is defined as the solution of the following equation: 
%\[\sum_{k=1}^K\left(\frac{f_0(z_k)}{f_0(z_k)+rf_1(z_k)}-\frac{rf_1(z_k)}{f_0(z_k)+rf_1(z_k)} \right)= 0\]
%where $z_1,...,z_K$ are sampled from $\pi^{mixt}_{r^*}$. The solution of this equation is found using a bisection method.  The optimal Bridge sampling estimator can also be computed solving an equation.
%However it  can also be obtained iteratively as explained in section \eqref{sec:intro},. In our   simulation study we use  this recursive scheme:
%\begin{equation*}
%    \hat{r}^{(t+1)} = \frac{\sum_{k=1}^K \frac{ f_0(z^1_k)}{  \hat{r}^{(t)}f_1(z^1_k)+f_0(z^1_k)}}{\sum_{k=1}^K \frac{ f_1(z^0_k)}{  \hat{r}^{(t)}f_1(z^0_k)+f_0(z^0_k)}}\qquad t=1,2,...
%\end{equation*}
%where $z_1^i,z_2^i,...,z_K^i$ are drawn from $p_i$ for $i \in \{0,1\}$. This procedure converges very fast in $t$. For every experiment, we run this recursion until convergence. }

As discussed in Section \ref{sec:stateoftheart}, the optimal Bridge sampling estimator is not available in a closed form, so we relied on the recursive algorithm described in equation \eqref{eq:optbridge}, which was run until convergence. For the RIS estimator, we used equation \eqref{eq:ris_est} using $\pi_{r^*}^{mixt}$ as proposal distribution.

We adjusted the sample sizes of each method to make sure that they are all comparable in terms of number of calls to functions $f_0$ and $f_1$. Four MH samplers were implemented. The first two are MH samplers whose invariant distributions are $p_0$ and $p_1$. These are used for BRIDGE-OPT, RIS-MIXT and SARIS-MIXT. Then two samplers generating non homogenous Markov chains with, at each step, invariant distributions being $\pi^{opt}_{r_k}$ and $\pi^{mixt}_{r_k}$, were built to compute respectively the SARIS-EXT-opt and the SARIS-EXT-mixt estimators. 

For each of the estimators, a budget of 2$K$+2$K_{heat}$ draws were allocated, with 2$K$ samples used to compute the estimators and 2$K_{heat}$ for heating the MH samplers. For the experiments, we used $K=5000$ and $K_{heat}=300$.

% For the two Bridge like ratio importance sampling estimators and the SARIS based on $\pi^{mixt}_{r^*}$ we considered a sample of $\tilde{K} = 2K$ using the same samplers used for Bridge sampling. For the estimator based on $\pi^{mixt}_r$ and $\pi^{opt}_r$ we used $2K_{heat}+2K$ iterations. For every SARIS procedure, the SARIS estimator is chosen as the mean of the sequence of estimated ratio over the last $K$ iterations.  \\

We also considered the estimation of $\log(r^*)$ using recursion \eqref{eq:satransfo}. From a theoretical point of view,  the use of this transformation is equivalent to using a delta method. Therefore, results obtained for the estimation of $\log r^*$ or $r^*$ are comparable in terms of performances. However, since the numerical stability of the procedure is greater in the former case, in the sequel we only present results associated with the estimation of $\log r^*$.

We first consider two cases: $1)$ a strong overlap between $p_0$ and $p_1$ with $\mu = 1$ and $2)$ a small overlap between $p_0$ and $p_1$ with $\mu=5$. Results are presented in Figure \ref{fig:compmu}. 

As the theory suggests, the optimal bridge sampling estimator outperforms the two other methods based on samples from $p_0$ and $p_1$. However, when considering the case $\mu=1$, the five methods present performances of the same order. The fact that the bridge like ratio importance sampling performs better than the SARIS-MIXT estimator can be interpreted by the fact that the ratio importance sampling estimator imposes at each step the estimator to solve the empirical version of the SARIS identity \eqref{eq:sa2}, which might enhance the stability of the procedure. This difference should be diminished by considering adaptive step sizes, to mimic the differences between  $\hat{r}^{ris}_{K}$ and $\hat{r}^{ris}_{K+1}$. \\ 

Figure \ref{fig:compmu} illustrates the fact that the SARIS-EXT estimators presented are much more robust to little overlap than ratio importance sampling and Bridge sampling. However, it also illustrates the fact that in more simple cases such as the one illustrated in Figure \ref{fig:compmu}, methods that only use draws from $p_0$ and $p_1$, which are easier to apply might be sufficient. 

\begin{figure}
    \centering
    \includegraphics[scale = 0.4]{ 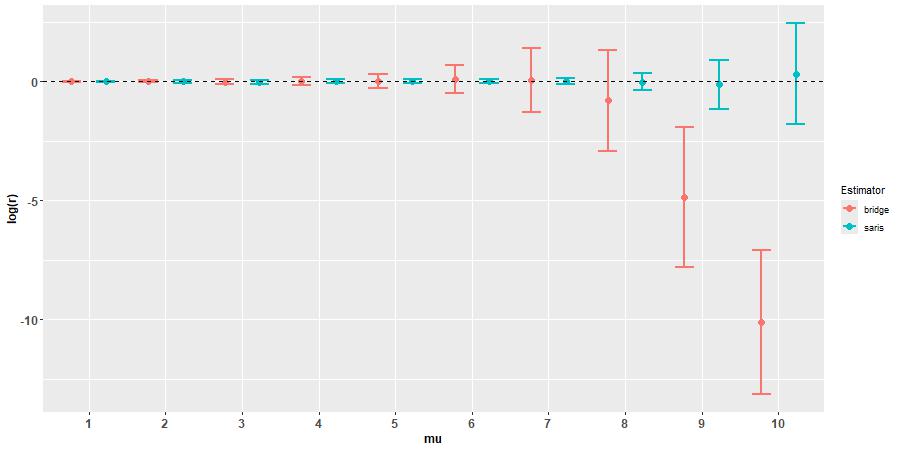}
    \caption{Estimation of the log ratio of normalizing constants of the densities of a $\mathcal{N}(0,1)$ and the one of a $\mathcal{N}(\mu,1)$ for varying values of $\mu$, using the optimal Bridge sampling estimator BRIDGE-OPT (red) and SARIS-EXT-opt to compute the ratio  The dots represent the empirical means and the error bars the empirical standard deviations computed over 50 repetitions.}
    \label{fig:muvar}
\end{figure}

To show intermediate results between the cases $\mu=1$ and $\mu=5$, and worse ones, we compare the optimal Bridge sampling estimator with the SARIS-EXT-opt estimator for varying values of $\mu$ between $1$ and $10$. For the simulations we use $K=5000$ samples for each expectation in the bridge sampling estimation procedure, and therefore use $2K$ samples for the SARIS procedure. Results are displayed in Figure \ref{fig:muvar}. For each value of $\mu$, 50 repetitions were computed. The dots represent the empirical means and the errorbars the empirical standard deviations computed over the 50 repetitions. Figure \ref{fig:muvar} illustrates the robustness of the proposed procedure in comparison to the bridge sampling estimator that highly deteriorates when the overlap reduces. This makes the proposed procedure appealing when little information is available on the distributions.% Indeed, as an example when considering the computation of a marginal likelihood, the \textit{bridgesampling} R package \cite{gronau2017bridgesampling} using a Gaussian approximation of the posterior distribution as proposal distribution, if this approximation is not good, the optimal bridge sampling estimator is likely to poorly perform. 

\subsection{Joint estimation in latent variables models}\label{sec:jointexp}

To illustrate the joint procedure presented above, we consider the following model of linear regression with missing values. Let $i=1,...,n$, we observe the response $y_i\in\mathbb{R}$ modeled as : 
\begin{equation}\label{eq:reglin}
    y_i = \beta_0+\beta_1x_{i1} +\beta_2x_{i2}+\varepsilon_{i}
\end{equation}
 
where $(\varepsilon_i)_{i=1,...,n}$ is a sequence of independent and identically distributed Gaussian noise with known variance $\sigma^2$, $\beta=(\beta_0,\beta_1,\beta_2)^T$ is an unknown vector of regression coefficients, and $(x_i)_{i=1,...,n}=(x_{i1},x_{i2})_{i=1,...,n}^T$ is an independent and identically distributed sample of covariates from a $\mathcal{N}\left((\mu_1,\mu_2)^T,\left(\begin{array}{cc}
     \gamma_1^2&0  \\
     0&\gamma_2^2 
\end{array}\right)\right)$. We suppose that for $i=1,...,r$ we only observe $(y_i,x_{i1})$ and for $i=r+1,...,n$ we observe $(y_i,x_{i1},x_{i2})$. This example is borrowed from the lecture notes of Julie Josse "Handling Missing values" available on this  \href{https://juliejosse.com/wp-content/uploads/2018/07/LectureMissing_Weij_modifAude.html}{website}\footnote{https://juliejosse.com/wp-content/uploads/2018/07/LectureMissing\_Weij\_modifAude.html}. Here the parameter to estimate is $\theta = (\beta,\gamma_1^2,\gamma_2^2,\mu_1,\mu_2)$. In order to make the notations as simple as possible, we will confound the notations of the random variables and their observed realizations. Furthermore, we are going to write $f_\theta(z)$ the density of the random variable $Z$ evaluated at $z$. For example $f_\theta(y_i|x_{i1})$ is the conditional density of $y_i$ given $x_{i1}$.

With these notations, the complete likelihood $L_n(\theta)$ is given by :

\begin{align*}
    L_n(\theta) &= \prod_{i=1}^n f_\theta(y_i,x_i)\\
    &= \prod_{i=1}^nf_\theta(y_i|x_i)f_\theta(x_i)
\end{align*}
However we do not observe the first $r$ $(x_{i2})_i$ that are handled as latent variables, therefore the observed likelihood is marginalized over their distribution : 

\begin{align*}
    L_n(\theta) &= \left[\prod_{i=1}^rf_\theta(y_i,x_{i1})\right] \times \prod_{i=r+1}^n f_\theta(y_i,x_i)\\
    &= \left[\prod_{i=1}^rf_\theta(y_i|x_{i1})f_\theta(x_{i1})\right] \times \prod_{i=r+1}^n f_\theta(y_i|x_i)f_\theta(x_i)\\
    &=\left[\prod_{i=1}^r\int_{x_{i2}}f_\theta(y_i|x_{i1},x_{i2})f_\theta(x_{i1})f_\theta(x_{i2}) dx_{i2} \right]\times \prod_{i=r+1}^n f_\theta(y_i|x_i)f_\theta(x_i)
\end{align*}

In fact we can compute exactly the marginal likelihood as the conditional distribution of $y_i$ given $x_{1i}$ is a $\mathcal{N}\left(\beta_1x_{i1}, \beta_2^2\gamma_2^2+\sigma^2\right)$. \\
Given a realization $\mathbf{x}_{2} = (x_{i2})_{i=1,..,r}$ of the unobserved variables, we introduce the complete log likelihood defined as : 

\[l_n(x_{12},...,x_{r2};\theta) = \sum_{i=1}^n\log\left(f_\theta(y_i,x_i)\right)\]

We consider here the following test : 

\begin{equation*}
    H_0:\quad\beta_0=0\quad against \quad H_1:\quad \beta_0\neq 0 
\end{equation*}

To apply the joint procedure described in Section \ref{sec:joint} we consider the unconstrained parameter space $\Theta_1$ corresponding to the alternative hypothesis, and the constrained parameter $\Theta_0$ that corresponds to the case $\beta_0=0$. For the two estimation procedures, we used stochastic gradient descent. At each step $k$ the procedure computes an estimator $g_k$ of  the log likelihood ratio,  jointly to the estimators $\hat\theta_{0,k}$ and $\hat\theta_{1,k}$ of respectively the restricted and unrestricted maximum likelihood estimators as follows : 
\begin{enumerate}
    \item Draw $\textbf{x}^{(k+1)}_{02}$ from the posterior distribution of $(x_{i2})_{i=1,...,r}$ given $((y_i,x_i)_{i=1,...,r},\hat\theta_{0k})$ and $\textbf{x}^{(k+1)}_{12}$ from the posterior distribution of $(x_{i2})_{i=1,...,r}$ given $((y_i,x_i)_{i=1,...,r},\hat\theta_{1k})$
    \item Update $\hat\theta_{0,k+1}$ and $\hat\theta_{1,k+1}$ each with a gradient step : 
    \begin{align*}
        \hat\theta_{0,k+1} &= \hat\theta_{0,k} - \gamma_k\nabla_\theta l_n(\textbf{x}^{(k+1)}_{02};\hat\theta_{0,k})\\
        \hat\theta_{1,k+1} &= \hat\theta_{1,k} - \gamma_k\nabla_\theta l_n(\textbf{x}^{(k+1)}_{12};\hat\theta_{1,k})
    \end{align*}
    \item Draw $\tilde{\textbf{x}}^{(k+1)}_{2}$ from a uniform distribution on $\{\textbf{x}^{(k+1)}_{02},\textbf{x}^{(k+1)}_{12}\}$
    \item Update $g_{k+1}$ as : 
    \[ g_{k+1} =g_k+ \gamma_k \frac{\exp\left\{l_n(\tilde{\textbf{x}}^{(k+1)}_2;\hat\theta_{0,k})\right\}
-\exp\left\{l_n(\tilde{\textbf{x}}^{(k+1)}_2;\hat\theta_{1,k})+g_k\right\}}{\exp\left\{l_n(\tilde{\textbf{x}}^{(k+1)}_2;\hat\theta_{0,k})\right\}
+\exp\left\{l_n(\tilde{\textbf{x}}^{(k+1)}_2;\hat\theta_{1,k})+g_k\right\}}   \]
\end{enumerate}

For the following experiments, we used a constant step size of $\gamma_k=0.1$ for the SARIS procedure and the estimation processes to keep the speed of convergence of the different iterations as similar as possible. The parameters used to generate the data are $\mu = (1,1)^T$, $(\gamma_1,\gamma_2) = (1,1)$, $\sigma^2=2$ and $\beta = (0.1,1,1)$. Finally, we considered $n=200$ individuals and two different numbers of missing values. We considered $r=20$ which corresponds to $10\%$ of missing values for the second covariate, and then $r=50$ which corresponds to $25\%$. We used $K=250$ iterations for the estimation process. We reproduced the experiment 20.

\begin{figure}[htbp]
    \centering
    \begin{minipage}[b]{0.45\textwidth}
        \centering
        \includegraphics[width=1.1\textwidth]{ 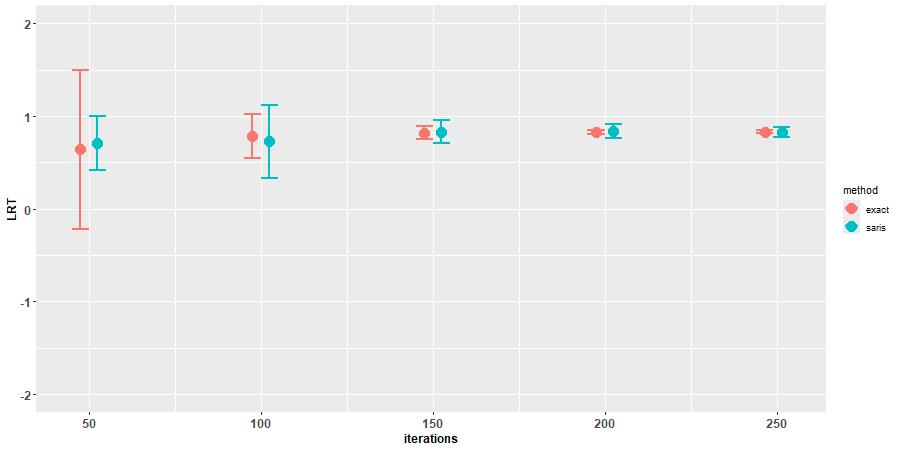}
    \end{minipage}
    \hspace{0.4cm}
    \begin{minipage}[b]{0.45\textwidth}
        \centering
        \includegraphics[width=1.1\textwidth]{ 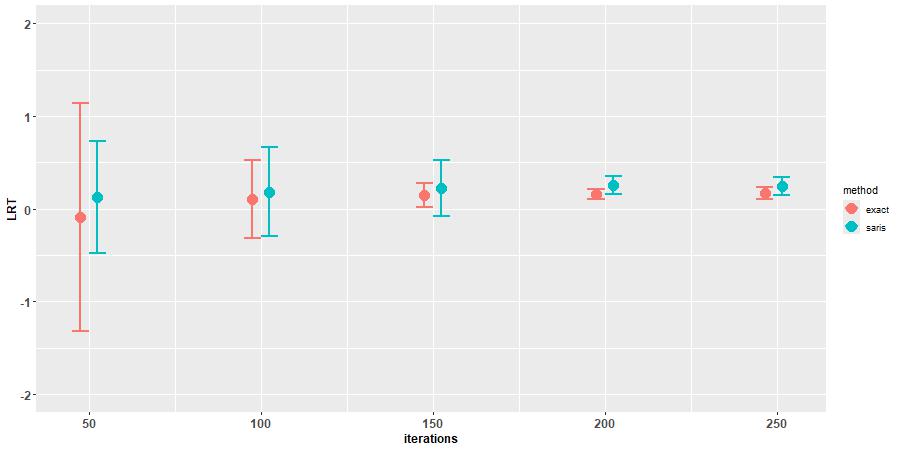}
    \end{minipage}
    \caption{{\footnotesize Estimation of the log likelihood in the  latent variable model of Section \ref{sec:jointexp} in the case $r=10\%n$ (on the left) and $r=25\%n$ (on the right). Comparison over the iterations of the exact likelihood ratio $\left(-2\log\frac{L_n(\hat\theta_{0,k})} {L_n(\hat\theta_{1,k})}\right)_{k\geq0}$ (\textit{exact} in red) and its approximation $\left(-2g_k\right)_{k\geq0}$ (\textit{approx} in blue) using the joint procedure described in Section \ref{sec:joint}.  The dots represents the means taken over 20 repetitions, the errorbars correspond to the empirical standard deviation.}}\label{fig:joint}
    \vspace{0.2cm}

\centering
    \begin{minipage}[b]{0.45\textwidth}
        \centering
        \includegraphics[width=1.1\textwidth]{ 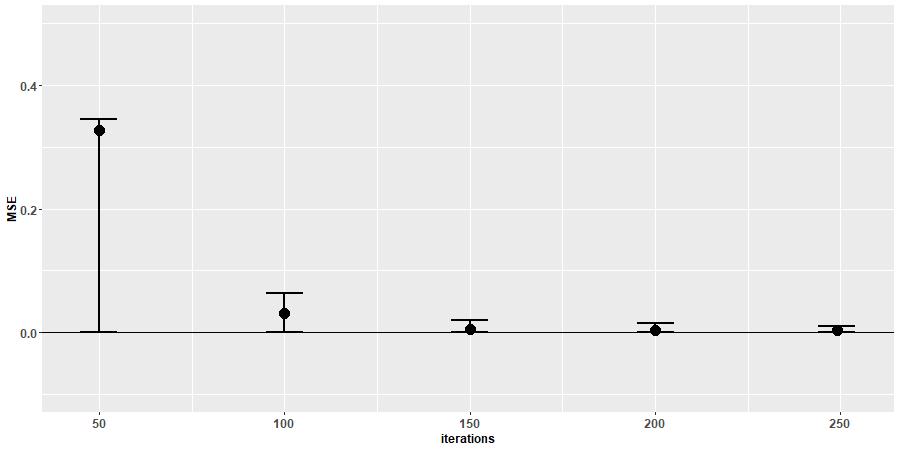}
    \end{minipage}
    \hspace{0.4cm}
    \begin{minipage}[b]{0.45\textwidth}
        \centering
        \includegraphics[width=1.1\textwidth]{ 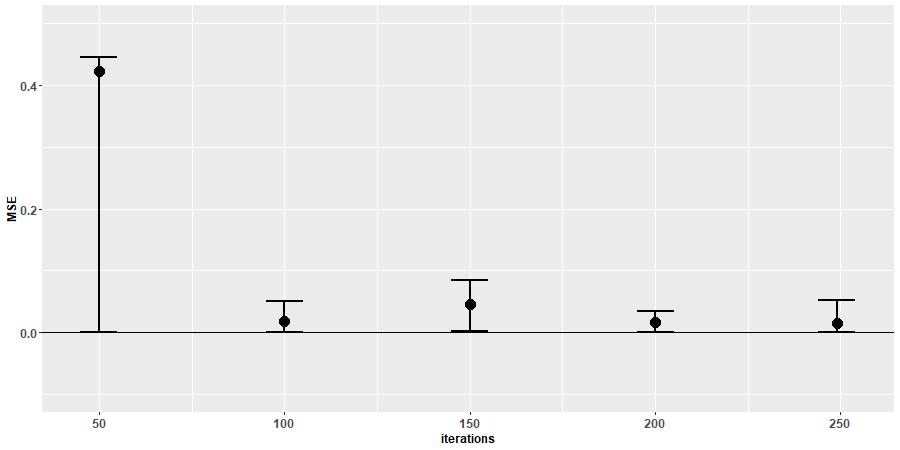}
    \end{minipage}
    \caption{{\footnotesize  Evolution of the empirical mean square error between the exact log likelihood ratio and its approximation using the joint procedure described in Section \ref{sec:joint} in latent variable model of Section \ref{sec:jointexp} in the case $r=10\%n$ (on the left) and $r=25\%n$ (on the right). The dots represents the means taken over 20 repetitions, the errorbars reach the 5\% and 95\% empirical quantiles.}}\label{fig:jointmse}

\end{figure}

Figure \ref{fig:joint} displays the evolution of the exact likelihood ratio $-2\log\frac{L_n(\hat\theta_{0,k})} {L_n(\hat\theta_{1,k})}$ over the iterations, compared to the one of the SARIS estimator obtained using the joint procedure described in Section \ref{sec:joint}. The dots represent the mean, and the errorbars the standard deviations computed over the 20 repetitions of the experiment.  Figure \ref{fig:jointmse} plots the mean squared error between the exact log likelihood ratio along the iteration process and its approximation using the SARIS joint procedure. The dots are the mean over the 20 repetitions, and the errorbars reach the 5\% and 95\% corresponding empirical quantiles. \\

In the case where $r=10\%n$  we observe that the joint procedure accurately tracks the exact value of the likelihood ratio along the estimation process which is very encouraging. We observe that in the more complex case where $r=25\%n$ the joint procedure correctly tracks the exact likelihood ratio statistic but with a small bias. There might be several sources for this bias such as the difference in convergence speed of both estimation processes or the higher variance due to the higher number of missing data. However, the approximation still tracks the exact value along iterations, which makes the proposed joined procedure appealing as a first step of a marginal likelihood computation. Indeed it gives a first estimator at the end of the estimation process that might be used as a starting point for a SARIS procedure, or an estimator $\hat{r}$ to use the approximated optimal scheme of \cite{chen1997monte} to use as proposal distribution $\pi\propto|f_0-\hat{r}f_1|$ in a ratio importance sampling procedure. \\

All the experiments were computed on R version 4.3.3 (2024-02-29 ucrt). For reproducibility the scripts are available at \href{https://github.com/tguedon/saris}{this link}\footnote{https://github.com/tguedon/saris}.  

\section{Conclusion}\label{sec:disc}

We proposed a new methodology to compute ratios of normalizing constants that relies on the principle of stochastic approximation. Our procedure presents good theoretical properties which makes it competitive with the best methods from the literature.  More precisely, our estimator is consistent and asymptotically Gaussian as the number of iterations goes to infinity. Moreover, the practical implementation of the algorithm reaches an  asymptotic variance which is smaller than the optimal variance of the Bridge sampling estimator. Another important advantage is that our estimator does not require to fix in advance the computational effort thanks to its iterative nature.  Indeed our procedure can be stopped in practice once a given convergence criterion is reached. Furthermore,  our estimator seems more robust to little overlap between the two unnormalised distributions considered and outperforms the Bridge sampling estimator in some of the numerical examples considered. The proposed methodology also allows for the computation of single marginal likelihoods. Moreover, in the context of likelihood ratio test statistics in latent variables models,  our  procedure can be integrated in the parameter estimation process to reduce the computational effort.

Besides these positive points, there are several interesting perspectives to investigate. Thanks to the rich literature on stochastic approximation and more specifically on stochastic gradient descent, many refinements can be explored, such as acceleration, variance reduction or adaptive step sizes.
Similarly, refinements used in classical Monte Carlo such as the Warp Bridge sampling \citep{meng2002warp,wang2022warp} could be applied to reduce the asymptotic variance of the estimator.  Furthermore, as a method to compute ratios of normalizing constants, the SARIS procedure can benefit from the use of intermediate distributions that enable to decompose the problem into several simpler sub-problems, in the principle of stepping stone sampling.

 \section{Declarations}
 \subsection{Funding}
 This work was funded by the Stat4Plant project ANR-20-CE45-0012.
\subsection{Conflict of interest}
The authors declare that they have no conflict of interest.

\bibliography{biblio}

\begin{appendices}
\section{Proofs}\label{secA1}
\subsection{Proof of Proposition \ref{prop:wp1 iid}}
We first rewrite the iterative scheme \eqref{eq:algoRM} for every positive integer $k$:
\[r_{k+1} = r_k + \gamma_{k+1}H_{\pi_{r_k}}(Z_{k+1}),\qquad Z_{k+1}\sim\pi_{r_k}
\]
introducing the notation  $H_{\pi_r}(z,r) = \frac{f_0(z)-rf_1(z)}{\pi_r(z)}$.

We apply the almost sure convergence theorem of  Robbins-Monro algorithms \citep{robbins1951stochastic} stated in  section 5.1 of \cite{benveniste2012adaptive} to prove Proposition \ref{prop:wp1 iid}. 

Under Assumption \ref{ass:density} we can define the function $h$ for every $r\in\bbr$ by $h(r) = \E_{r}\left[H_{\pi_r}(Z,r)\right] $ where $\E_{r}$ stands for $\E_{\pi_r}$ for sake of simplicity.  It follows directly that $h(r) = c_0-rc_1$ for every $r\in\bbr$ and $h(r^*)=0$.  We  define the filtration $(\mathcal{F}_k)_{\geq0}$ corresponding to the increasing family of $\sigma$-algebra generated  by $(r_0,Z_1,....,Z_k)$.  We now verify the following assumptions of the theorem stated in section 5.1 of \cite{benveniste2012adaptive}:
\begin{enumerate} 
\item for any positive measurable function $g$ defined on $\mathcal{Z} \times \mathbb{R}$ we have $ \E\left[ g(Z_{k+1},r_{k}) |\mathcal{F}_k\right] = \E_{r_{k}}\left[g(Z,r_k)\right] .$
\item there exists $C>0$, such that for every $r\in \bbr$ $\E_{r}\left[H_{\pi_r}(Z,r)^2\right]\leq C(1+r^2)$
\item there exists $r^*>0$ such that for every $r\in\bbr\setminus\{r^*\}$:
\[(r-r^*)h(r)<0\]
\item $\sum \gamma_k=+\infty$ and $\sum \gamma_k^2<+\infty$
\end{enumerate}
%\begin{remark}
%    The third point is even stronger than what is required in \cite{benveniste2012adaptive}, but it is not a problem in our setting, and it lightens the notations. 
%\end{remark}

The first point is straightforward considering the sampling scheme of equation \eqref{eq:algoRM}. \\
We now consider point 2):
\begin{align*}
    \E_{r}\left[H_{\pi_r}(Z,r)^2\right]&= \int_{\mathcal{Z}}\left( \frac{f_0(z)-r f_1(z)}{\pi_r(z)}\right)^2\pi_r(z)\mu(dz)\\
    &=\int_{\mathcal{Z}}\frac{\left(f_0(z)-r f_1(z)\right)^2}{\pi_r(z)}\mu(dz) \\
    &= \int_{\mathcal{Z}} | H_{\pi_r}(z,r)|  | f_0(z)-r f_1(z)| \mu(dz) \\
    &\leq \int_{\mathcal{Z}}|H_{\pi_r}(z,r)|f_0(z)\mu(dz)+ \int_{\mathcal{Z}}|H_{\pi_r}(z,r)|f_1(z) | r | \mu(dz)\\
    &\leq c_0\E_0\left[|H_{\pi_r}(z,r)|\right] +  | r |  c_1\E_1\left[|H_{\pi_r}(z,r)|\right]\\
  %  &\leq c_0\E_0\left[|H_\pi(z,r)|\right] + r^2 \frac{c_1}{r}\E_1\left[|H_\pi(z,r)|\right]\\
  %  &\leq C(1+r)\\
    &\leq \tilde{C}(1+r^2)
\end{align*}
 thanks to Assumption \ref{ass:density}.
 
Point 3) is straightforward as $h(r)=c_1(r^*-r)$. Finally point 4) is implied by  Assumption \ref{ass:stepsize2}.\hfill$\square$

\subsection{Proof of Proposition \ref{prop:tcl}}
We apply the central theorem for Robbins-Monro algorithms (\cite{duflo1996algorithmes} chapter 4) to prove Proposition \ref{prop:tcl}. The additional assumptions to verify are: 
\begin{enumerate}
    \item There exists a neighborhood $U$ of $r^*$ such that the function $h$  is continuously differentiable on $U$ and  for all $r\in U$, $h'(r)<0$ .
    \item There exists $\Gamma>0$, such that almost surely: \[\underset{k\rightarrow+\infty}{\lim}\E\left[\left(H_{\pi_{r_k}}(Z_{k+1},r_k)-h(r_k)\right)^2|\mathcal{F}_k\right] = \Gamma\]
    \item There exists $\delta>0$ such that \[\underset{k}{\sup}\quad\E\left[\left(H_{\pi_{r_k}}(Z_{k+1},r_k)-h(r_k)\right)^{2+\delta 
 }|\mathcal{F}_k\right]<+\infty\]
 \item There exist  $\frac{1}{2}<\epsilon<1$, $a>0$, $b>0$ such that the sequence of step sizes $(\gamma_k)$ is of the form  $\gamma_k=\frac{a}{b+k^\epsilon}$.
\end{enumerate}
%under this assumptions, the averaged sequence $r_k^{AV}$ is asymptotically Gaussian with asymptotic variance $V$ defined as follows: 
%\begin{equation}\label{eq:tclAPP}V = \frac{\Gamma}{h'(r^*)^2}\end{equation}

The first point is straightforward as $h$ is linear in $r$ and $h'$ is constant equal to $-c_1<0$. To prove the second point, let us introduce $\xi_{k+1}=H_{\pi_{r_k}}(Z_{k+1},r_k)-h(r_k)$ for all integer $k$. After some calculation, we get:
$$\E(\xi_{k+1 }^2|\mathcal{F}_k)=  c_0\E_0\left[H_{\pi_{r_k}}(Z,r_k)\right]- r_kc_1\E_1\left[H_{\pi_{r_k}}(Z,r_k)\right]-h(r_k)^2$$

%  we can note that for every $r\in\bbr$:
% \begin{align*}
%     \E_{r}\left[\left(H_\pi(z,r)-h(r)\right)^2\right]&= E_{\pi(.,r)}\left[H_\pi(z,r)^2\right] - h(r)^2\\
%     &= \int \left(\frac{f_0(z)-f_1(z)r}{\pi(z)}\right)^2\pi(z;r)\mu(dz) \\
%     &= \int \frac{\left(f_0(z)-f_1(z)r\right)^2}{\pi(z;r)}\mu(dz) \\
%     &= \int \frac{f_0(z)-f_1(z)r}{\pi(z;r)}\left(f_0(z)-f_1(z)r\right)\mu(dz) \\
%     &= c_0\E_0\left[H_\pi(z,r)\right]- rc_1\E_1\left[H_\pi(z,r)\right]
% \end{align*}
%therefore,
%  \begin{align*}
%     \E\left[\left(H_\pi(Z,r_k)-h(r_k)\right)^2|\mathcal{F}_k\right]= c_0\E_0\left[H_\pi(Z,r_k)\right]- r_kc_1\E_1\left[H_\pi(Z,r_k)\right]
% \end{align*}

Under Assumptions \ref{ass:density} and \ref{ass:stepsizetcl},  the sequence $(r_k)_k$ converges almost surely to $r^*$, $h(r^*)=0$,  and,  applying the dominated convergence theorem,  the following convergence holds almost surely: 
\[\underset{k\rightarrow+\infty}{\lim}\quad \E(\xi_{k+1 }^2|\mathcal{F}_k)=\E_{r^*}\left[H_{\pi_r^*}(Z,r^*)^2\right] \]
The right hand side term  is positive under Assumption \ref{ass:nonprop}.
%\[\underset{k\rightarrow+\infty}{\lim}\quad \E_{r_k}\left[\left(H_{\pi_{r^*}}(Z,r_k)-h(r_k)\right)^2\right]=\E_{r^*}\left[H_\pi(Z,r^*)^2\right] = \Gamma\]
 Finally we show that Assumption \ref{ass:regtcl} implies condition 3)  using similar calculation and arguments.  Assumption \ref{ass:stepsizetcl} corresponds to condition 4).
 
% It is straightforward to notice that for every $r\in\bbr$:
% \begin{align*}
%     \E_{r}\left[H_\pi(z,r)^{2+\delta}\right]&= \int\left(\frac{f_0(z)-f_1(z)r}{\pi(z;r)}\right)^{2+\delta}\pi(z;r)\mu(dz)\\
%     &= \int \left(\frac{f_0(z)-f_1(z)r}{\pi(z;r)}\right)^{1+\delta}\left(f_0(z)-f_1(z)r\right)\mu(dz) \\
%     &= c_0\E_0\left[H_\pi(z,r)^{1+\delta}\right]- rc_1\E_1\left[H_\pi(z,r)^{1+\delta}\right]
% \end{align*}
 
 Applying the central theorem for Robbins-Monro algorithms (\cite{duflo1996algorithmes} chapter 4) we get that:
   $$V_{saris}(\pi_{r^*}) =\E_{r^*}(H_{\pi_{r^*}}(Z,r^*)^2) /c_1^{2}   $$

We then apply Jensen  inequality to get the minorization:
\begin{align*}
  V_{saris}(\pi_{r^*})     &\geq \left(\E_{r^*}\left[|H_{\pi_{r^*}}(Z,r^*)|\right] \right)^2  /c_1^{2}\\
    &= \left( \int_{\mathcal{Z}}|f_0(z)- r^*  f_1(z)| \mu(dz)\right)^2  /c_1^{2} \\
    &= {r^*}^2\left( \int_{\mathcal{Z}}|p_0(z)-p_1(z)|\mu(dz)\right)^2
\end{align*}
Moreover the equality case holds in Jensen inequality for $\pi^{opt}_{saris}(z) \propto |f_0(z)-r^*f_1(z)|$ which leads to the result. \hfill$\square$

\subsection{Proof of Proposition \ref{prop:sarisextcvps}}
This proof follows the lines of the proof of Proposition \ref{prop:wp1 iid}. The only difference is that the objective function $h$ to nullify is : 

\[h:r\mapsto   \frac{c_0-rc_1}{c(r)}\] 
that verifies for every $r\in\bbr,$ $(r-r^*)h(r) = \frac{-c_1(r-r^*)^2}{c(r)}<0$. The remaining hypothesis are verified thanks to Assumption \ref{ass:supincrementtilde}. 

\begin{remark}
    The fact that $c(r)$ is strictly positive for all $r$ is implied by Assumption \ref{ass:supincrementtilde}. $c(r) = 0$ would mean that $\tilde{\pi}_r(Z)=0$ $\mu-$almost surely. 
\end{remark}
\hfill$\square$

\subsection{Proof of Proposition \ref{prop:sarisexttcl}}

In order to prove this proposition we proceed in two steps. The first step establishes the central limit theorem, the second step shows that the use of the unnormalized density $\tilde\pi^{opt}_r(z) = |f_0(z) - rf_1(z)|$  enables to reach the optimal asymptotic variance $V_{saris}^{opt}$. \\
The first step follows the same lines as the proof of Proposition \ref{prop:tcl}, replacing Assumptions \ref{ass:supincrement} and \ref{ass:regtcl} by Assumptions \ref{ass:supincrementtilde} and \ref{ass:regtcltilde}. Assumption \ref{ass:c} ensures the differentiability of $h$ in a neighborhood $U$ of $r^*$, and the fact that $h'(r^*)\neq 0$. Furthermore, $h'(r^*) = \frac{-c_0c(r^*) - c'(r^*)(c_0-r^*c_1)}{c(r^*)^2} = \frac{-c_1}{c(r^*)}$. The continuity of $c$ and $r\mapsto \tilde\pi_r(z)$ for every $z$ are guaranteed by Assumption \ref{ass:density} and the differentiability of $h$. The central limit theorem of \cite{duflo1996algorithmes} chapter 4 concludes this step, and :
\begin{align*}
    V_{ext}(\tilde\pi_{r^*}) &= \frac{c(r^*)^2}{c_1^2}\E_{r^*}\left[\left(\frac{f_0(Z)-r^*f_1(Z)}{\tilde\pi_{r^*}(Z)}\right)^2\right]\\ &= \frac{1}{c_1^2}\E_{r^*}\left[\left(\frac{f_0(Z)-r^*f_1(Z)}{\pi_{r^*}(Z)}\right)^2\right] \\&= V_{saris}(\pi_{r^*})
\end{align*}

Let $r\in\bbr$, $\tilde\pi^{opt}_r:z\mapsto  |f_0(z) - rf_1(z)|$ and $c^{opt}:r\mapsto  \int_\mathcal{Z}\tilde\pi^{opt}_r(z)dz$. The second step proves that $\tilde\pi^{opt}$ verifies Assumptions \ref{ass:supincrementtilde}, \ref{ass:regtcltilde} and \ref{ass:c}. \\
First for every $r\in\bbr$ and $z\in\mathcal{Z}$, the quantity $\frac{f_0(z)-rf_1(z)}{\tilde\pi^{opt}_r(z)}$ is bounded since its absolute value equals 1, and therefore Assumptions \ref{ass:supincrementtilde} and \ref{ass:regtcltilde} are verified. The main point to verify is Assumption \ref{ass:c}. This is not straightforward because of the absolute value in the definition of the $c^{opt}$ function. Let $r\in\bbr$,  $\mathcal{A}_r^+ = \{z\in\mathcal{Z}: f_0(z)> rf_1(z)\}$ and  $\mathcal{A}_r^- = \{z\in\mathcal{Z}: f_0(z)< rf_1(z)\}$. We get
\begin{align*}
    c(r) &= \int_\mathcal{Z}|f_0(z)-rf_1(z)|\mu(dz) \\
    &= -\int_{\mathcal{A}_r^-}\left(f_0(z)-rf_1(z)\right)\mu(dz) + \int_{\mathcal{A}_r^+}\left(f_0(z)-rf_1(z)\right)\mu(dz) \\
    &= -c_0+ r c_1 + 2 \int_{\mathcal{A}_r^+}\left(f_0(z)-rf_1(z)\right)\mu(dz)
\end{align*}
using Assumption \ref{ass:piopt}, and that for $i=0,1$, $c_i = \int_{\mathcal{A}_r^+}f_i(z)\mu(dz) +\int_{\mathcal{A}_r^-}f_i(z)\mu(dz)$. \\
To study the differentiability of  $m:r\mapsto \int_{\mathcal{A}_r^+}f_0(z)-rf_1(z)\mu(dz)$, we consider $r_0\in\bbr$ and $\epsilon>0$:
\begin{align*}
    m(r_0+\epsilon)-m(r_0)&= \int_{\mathcal{A}_r^+\backslash \mathcal{A}_{r+\epsilon}^+} \left(rf_1(z)-f_0(z)\right)\mu(dz) - \epsilon \int_{\mathcal{A}_{r+\epsilon}^+}f_1(z)\mu(dz) 
\end{align*}

Let study the first term. 
\begin{align*}
    \mathcal{A}_r^+\backslash \mathcal{A}_{r+\epsilon}^+ &= \{z\in\mathcal{Z}: rf_1(z)<f_0(z)\leq (r+\epsilon)f_1(z)\} \\
    &=\{z\in\mathcal{Z}: 0<f_0(z) - rf_1(z)\leq \epsilon f_1(z)\} 
\end{align*}

$f_1$ is $\mu-$almost surely upper-bounded  as it is a (unnormalized) density, let $\bar{f}_1$ its upperbound. 

\begin{align*}
    |\int_{\mathcal{A}_r^+\backslash \mathcal{A}_{r+\epsilon}^+} rf_1(z)-f_0(z)\mu(dz)|&\leq \int_{\mathcal{A}_r^+\backslash \mathcal{A}_{r+\epsilon}^+} |rf_1(z)-f_0(z)|\mu(dz) \\
    &\leq \epsilon\bar{f}_1\mu(\mathcal{A}_r^+\backslash \mathcal{A}_{r+\epsilon}^+)
\end{align*}

and $\mu(\mathcal{A}_r^+\backslash \mathcal{A}_{r+\epsilon}^+)\underset{\epsilon\rightarrow 0}{\rightarrow}0$ with Assumption \ref{ass:piopt}. The same calculations hold with $\epsilon<0$ and we get that $m$ is differentiable on $\bbr$ and for every $r\in\bbr$ $m'(r) = -\int_{\mathcal{A}_r^+}f_1(z)\mu(dz)$ and finally $c$ is differentiable on $\bbr$ and specifically : 
\begin{align*}
    c'(r^*) &= c_1 - 2\int_{\mathcal{A}_{r^*}^+}f_1(z)\mu(dz)
\end{align*}

therefore $h$ is differentiable on $\bbr$ which concludes the proof, as the first part of the proposition shows that  $V_{ext}(\tilde\pi_{r^*}) = V_{saris}(\pi_{r^*})$. 

\subsection{Proof of Proposition \ref{prop:samixtcomp}}
We first derive the expression of $ V_{ext}(\tilde\pi^{mixt}_{r^*}) $. 

We recall the notations: $\pi^{mixt}_{r^*}(z)= \frac{1}{2}\left(p_0(z)+p_1(z)\right)\propto f_0(z)+r^*f_1(z) = \tilde\pi^{mixt}_{r^*}(z)$. 

We start from the result of Proposition \ref{prop:tcl}:

 \begin{align*}
    V_{ext}(\tilde\pi^{mixt}_{r^*})  &=  \int_\mathcal{Z}\frac{\left(f_0( z)-r^*f_1( z)\right)^2}{\pi^{mixt}_{r^*}( z)} \mu(dz) / c_1^{2}\\
     &= 2{r^*}^2\int_\mathcal{Z}\frac{\left(p_0( z)-p_1( z)\right)^2}{p_0( z)+p_1( z)} \mu(dz)  \\
   %  &= 2{r^*}^2\left(  2 + \int_\mathcal{Z}-(p_0( z)+p_1( z))+\frac{\left\{p_0( z)-p_1( z)\right\}^2}{p_0( z)+p_1( z)} \mu(dz)  \right) \\
  %   &= 2{r^*}^2\left(  2 + \int_\mathcal{Z}\frac{\left\{p_0( z)-p_1( z)\right\}^2-(p_0( z)+p_1( z))^2}{p_0( z)+p_1( z)} \mu(dz)  \right) \\
  %   &= 2{r^*}^2\left(  2 + \int_\mathcal{Z}\frac{\left\{p_0( z)-p_1( z) + p_0( z)+p_1( z)\right\}\left\{p_0( z)-p_1( z) - p_0( z)-p_1( z)\right\}}{p_0( z)+p_1( z)} \mu(dz) \right) \\
   %   &= 2{r^*}^2\left(  2 + \int_\mathcal{Z}\frac{2p_0( z)\times(-2p_1( z))}{p_0( z)+p_1( z)}d z \right) \\
    &= 2{r^*}^2\left(   \int_\mathcal{Z}\frac{\left(p_0( z)+p_1( z)\right)^2-4p_0( z)p_1( z)}{p_0( z)+p_1( z)} \mu(dz)  \right) \\
      &= 2{r^*}^2\left(  2 - \int_\mathcal{Z}\frac{4p_0( z)p_1( z)}{p_0( z)+p_1( z)} \mu(dz)  \right) \\
      &=4{r^*}^2\left(  1 - \Psi \right)
 \end{align*}
using $(a-b)^2=(a+b)^2-4ab$ for all  $a$ and $b$ reals. The result of Proposition \ref{prop:sarisexttcl} also holds as $h(r)=\frac{c_0-rc1}{c_0+rc_1}$ verifies Assumption \ref{ass:c}. 

We then present the main steps to prove the almost sure convergence and the asymptotic normality of the sequence obtained using Algorithm \ref{alg:sarismixt}. The analysis of this procedure follows the same lines as the one of the SARIS and SARIS-ext procedure. We introduce the notation  
\[\tilde{H}(z,r)=\frac{f_0(z)-rf_1(z)}{f_0(z)+rf_1(z)}\] and we define under Assumption \ref{ass:density}, with $\pi=\pi^{mixt}$, the function 
\[
	\tilde{h}(r)=\E_{\pi^{mixt}_{r^*}}\left(\frac{f_0(Z)-rf_1(Z)}{f_0(Z)+rf_1(Z)}\right).
\] 

After some calculation, we get 
$$\tilde{h}'(r^*)=- \frac{\Psi}{2r^*}$$ 
and 
$$\tilde{\Gamma}=\E_{\pi^{mixt}_{r^*}}\left[\left(\frac{f_0(Z)-r^*f_1(Z)}{f_0(Z)+r^*f_1(Z)}\right)^2\right] = (1-\Psi),$$ 
which is positive under Assumption \ref{ass:nonprop}.  Following the same lines as in the proofs of Propositions  \ref{prop:wp1 iid}  and \ref{prop:tcl}, it is straightforward to check that all the assumptions required for the consistency and asymptotic normality hold. Applying these results to the sequence generated by Algorithm  \ref{alg:sarismixt}  leads to the following asymptotic variance $V_{saris}^{mixt}$: 
\[V_{saris}^{mixt} =4 {r^*}^{2} (1-\Psi)/ \Psi^2 \] 

The optimal bridge sampling asymptotic variance is given in equation \eqref{eq:var_opt_bs} by $V^{opt}_{bridge} = 4 {r^*}^{2} (\Psi^{-1} - 1)$ and $V_{ris}^{mixt} = 4 {r^*}^{2} (1-\Psi)/\Psi^2$ (see for example \citet[Theorem 5.2.]{chen1997estimating}), which concludes the proof.

\hfill$\square$

%TROP DETAILLE
%In order to derive $V^{mixt}$ one should note that $h'$ is no longer constant. Indeed, for every $r\in\bbr$:
%\[h(r) = \int_{\mathcal{Z}}\frac{f_0(z)-f_1(z)r}{\pi^{mixt}(z,r)}\pi^{mixt}(z,r^*)\mu(dz)\]
%Let $U$ a neighborhood of $r^*$. For every $z\in\mathcal{Z}$, and $r\in U \cup \bbr_+^*$: 
%\[\left|\frac{f_0(z)-f_1(z)r}{f_0(z)+rf_1(z)}\right|\leq1\]
%therefore , for every $r\in U \cup \bbr_+^*$:
%\begin{align*}
%    h'(r)&= \int_{\mathcal{Z}} \frac{\partial}{\partial r}\left(\frac{f_0(z)-f_1(z)r}{\pi^{mixt}(z,r)}\right)\pi^{mixt}(z,r^*)\mu(dz)\\
%    &= \int_{\mathcal{Z}} \frac{\partial}{\partial r}\left(\frac{f_0(z)-f_1(z)r}{f_0(z)+rf_1(z)r}\times(c_0+rc_1)\right)\pi^{mixt}(z,r^*)\mu(dz)\\
%    &= -2 \int_{\mathcal{Z}}\frac{f_0(z)f_1(z)}{(f_0(z)+rf_1(z))^2}\pi^{mixt}(z,r^*)\mu(dz)(c_0+rc_1)+c_1\int_{\mathcal{Z}}\frac{f_0(z)-rf_1(z)}{f_0(z)+rf_1(z)}\pi^{mixt}(z,r^*)\mu(dz)\\
%\end{align*}
%evaluating it at $r=r^*$, noticing that that the second terms cancels as $c_0-r^*c_1=c_0-c_0=0$, we have: 
%\[h'(r^*)=-2\int_{\mathcal{Z}}\frac{f_0(z)f_1(z)}{f_0(z)+r^*f_1(z)}\mu(dz) = -c_1\Psi\]

 \end{appendices}

\end{document}